\documentclass[12pt]{article}

\AtBeginDocument{
  \addtocontents{toc}{\normalsize}
}

\pdfoutput=1 

\usepackage{amsmath,amssymb,amsfonts,amsthm,amscd,mathrsfs}
\usepackage{xcolor}
\definecolor{darkblue}{rgb}{0.1,0.1,.7}
\usepackage[colorlinks, linkcolor=darkblue, citecolor=darkblue, urlcolor=darkblue, linktocpage]{hyperref} 
\usepackage[square, comma, compress,numbers]{natbib}
\usepackage[]{version}
\usepackage[]{graphicx}
\usepackage[]{latexsym}
\usepackage{geometry}
\usepackage[all,cmtip]{xy}
\usepackage[margin=10pt,font=small,labelfont=bf]{caption}
\usepackage{ifthen}
\usepackage{tikz}
\usepackage{array,setspace,mathrsfs,amsfonts,yfonts,dsfont,bbm,colonequals}
\usepackage{dsfont} 
\usepackage{xspace}
\usepackage{subcaption}
\usepackage[normalem]{ulem}

\usepackage{accents}
\newlength{\dhatheight}

\def\blue{\color [rgb]{0.1,0.1,0.9}}

\newcommand{\reef}[1]{(\ref{#1})}

\def\eps{\epsilon}
\newcommand{\beq}{\begin{equation}} 
\newcommand{\eeq}{\end{equation}}
 
\def\nn{\nonumber} 
 
\def\bR {\mathbb{R}}

\def\calO {{\cal O}} 
\def\calG {{\cal G}} 
 
\def\calF {{\cal F}}

\def\half{{\textstyle\frac 12}}

\def\ge{\geqslant}
\def\le{\leqslant}

\def\nn{\nonumber}

\def\eps{\epsilon}

\newcommand{\be}{\begin{equation}}
\newcommand{\ee}{\end{equation}}
\def\ba{\begin{array}}
\def\ea{\end{array}}

\newcommand{\D}{\Delta}
\newcommand{\Df}{{\Delta_\phi}}

\numberwithin{equation}{section}
\begin{document}

\vspace*{-.6in} \thispagestyle{empty}
\begin{flushright}
CERN TH/2017-176
\end{flushright}
\vspace{1cm} {\large
\begin{center}
{\bf A tauberian theorem for the conformal bootstrap}
\end{center}}
\vspace{1cm}
\begin{center}
{\bf Jiaxin Qiao$^{a,b}$ and Slava Rychkov$^{a,b}$}\\[2cm] 
{
\small
$^a$  CERN, Theoretical Physics Department, 1211 Geneva 23, Switzerland\\
$^b$ Laboratoire de physique th\'eorique,\\ D\'epartement de physique de l'ENS,
\'Ecole normale sup\'erieure, PSL Research University,\\ 
Sorbonne Universit\'es, UPMC Univ.~Paris 06, CNRS, 75005 Paris, France
\normalsize
}
\end{center}

\vspace{4mm}
\begin{abstract}
For expansions in one-dimensional conformal blocks, we provide a rigorous link between the asymptotics of the spectral density of exchanged primaries and the leading singularity in the crossed channel. Our result has a direct application to systems of ${\rm SL}(2,\bR)$-invariant correlators (also known as 1d CFTs). It also puts on solid ground a part of the lightcone bootstrap analysis of the spectrum of operators of high spin and bounded twist in CFTs in $d>2$.
In addition, a similar argument controls the spectral density asymptotics in large $N$ gauge theories.
\end{abstract}

\vspace{1cm}
\hspace{0.3cm} \noindent August 2017

\newpage

{
\tableofcontents
}


\setlength{\parskip}{0.04in}

\section{Introduction}
In conformal field theory (CFT), channel duality relates the low- and high-dimension parts of the operator spectrum. The best-known example is Cardy's formula for the asymptotic density of states in 2d CFT \cite{Cardy:1986ie}, which follows from the modular invariance of the torus partition function. In this paper we will be concerned with another type of channel duality---the crossing symmetry of the four-point (4pt) function, which is the condition which forms the basis of the conformal bootstrap \cite{Ferrara:1973yt,Polyakov:1974gs,Mack:1975jr,Belavin:1984vu,Rattazzi:2008pe}. The 4pt function allows a convergent expansion in one channel, which should agree with the operator product expansion (OPE) in another channel. This was used in \cite{Pappadopulo:2012jk,Rychkov:2015lca} to put bounds on the spectral density of high-dimension operators, weighted by their OPE coefficients.
In \cite{Fitzpatrick:2012yx,Komargodski:2012ek} and many subsequent works,\footnote{See e.g.~\cite{Kaviraj:2015cxa, Alday:2015eya,Kaviraj:2015xsa,Alday:2015ota,Alday:2015ewa,Li:2015itl,Simmons-Duffin:2016wlq}.} similar constraints near the Minkowski lightcone were used to study asymptotics of operators with high spin and fixed twist.

In this paper we will revisit the problem of extracting large dimension asymptotics from crossing symmetry. Our goal will be to put on more solid ground some intuitive assumptions made in the previous work. For conceptual clarity, we will focus on an analogous problem in the context of CFTs in $d=1$. We will comment on the relevance of our work to $d>1$ in appendix \ref{d>1}.

\section{Formulation of the problem and the main result}
\label{formulation}

In a 1d unitary CFT,\footnote{See appendix \ref{CFT1} for what we mean by this.} consider 4pt function of a hermitean operator $\phi$ of scaling dimension $\Delta_\phi>0$:
\beq
\langle \phi(0)\phi(z)\phi(1)\phi(\infty)\rangle = z^{-2\Delta_\phi} \calG(z)\,.
\eeq
The function $\calG(z)$ has a decomposition in conformal blocks \cite{DO1,DO2,DO3}:
\beq
\calG(z) = \int_0^\infty d\Delta\, p(\Delta)\, G_{\Delta}(z),\qquad G_\D(z)=z^\D {}_2 F_1(\D,\D,2\D;z)\,.
\label{CBdec}
\eeq
Here $p(\Delta)$ is a non-negative spectral density. In the discrete spectrum case it's a sum of delta-functions with positive coefficients. Our arguments below will apply to both discrete and continuous spectrum case. A generalization to the case of unequal external dimensions will be considered in section \ref{unequal}.
 
We will only consider the 4pt function for real $0<z<1$. Conformal block decomposition \reef{CBdec} converges on this interval \cite{Pappadopulo:2012jk}. In fact the 4pt function can be analytically continued to complex $z$ (see e.g.~a recent discussion in \cite{Rychkov:2017tpc}), but we will not make use of this fact.

Consider the 4pt function in the limit $z\to 1$. This limit is dominated by the unit operator in the OPE $\phi(z)\phi(1)$. We will assume the unit operator is separated by a positive gap from the rest of the spectrum. In this case we have the asymptotics:\footnote{We use $A\sim B$ to mean that $A/B\to 1$ in the appropriate limit.}
\beq
\calG(1-x)\sim x^{-2\Delta_\phi} \qquad (x\to 0)\,.
\label{zto1}
\eeq
What can we say about the spectral density of the conformal block expansion \reef{CBdec} from the fact that it should give rise to such a powerlaw? This was first discussed in \cite{Pappadopulo:2012jk}, where an upper bound on the spectral density was given, which was then used to control the rate of convergence of the conformal block expansion. Subsequently, Ref.~\cite{Fitzpatrick:2012yx} addressed a more nuanced question about the \emph{asymptotics} of the spectral density, as opposed to just an upper bound.\footnote{More precisely, \cite{Fitzpatrick:2012yx,Komargodski:2012ek} studied the operators of high spin and constant twist in CFTs in $d>1$ dimensions, analyzing the bootstrap equation near the Minkowski lightcone. In doing so they encountered a problem formally equivalent to the one we are discussing. See appendix \ref{d>1}. Ref.~\cite{Komargodski:2012ek} did not discuss specifically the spectral density, so that we will be primarily comparing with \cite{Fitzpatrick:2012yx}.} They determined the asymptotics via the following argument.

Since the individual conformal blocks only have a mild $\log(1-z)$ singularity, the powerlaw singularity can appear only as a cumulative effect from the tail of the distribution at $\Delta \gg 1$. In this region, and for $z\to 1$, one can approximate conformal blocks as
\beq
\label{CBappr}
G_\Delta(1-x)\approx C(\D)\,K_0(2\sqrt{x}\D),\qquad C(\D)={4^\D \sqrt{\D/\pi}} \,.
\eeq
This approximation is valid for $\D\gg1$, $x\ll 1$, $x\D\ll 1$. So one concludes
\be
\label{Gappr}
\int_{\Delta_0}^\infty\, d\Delta\, C(\Delta) p(\Delta)  K_0(2\sqrt{x}\D) \sim x^{-2\Delta_\phi}\qquad (x\to 0).
\ee
The lower limit of the integral is unimportant since it's large $\D$ which dominate the asymptotics, so we set it to some fixed $\Delta_0$.

Now, a natural way to get the integral in \reef{Gappr} to scale as a powerlaw with $x$ is to assume that $C(\Delta)  p(\Delta)$ itself scales asymptotically as a powerlaw with $\Delta$. Taking a general powerlaw parametrization
\beq
\label{ptilde0}
C(\Delta) p(\Delta) \sim A^{-1} \Delta^{\gamma-1}\qquad(\Delta\to\infty)\,,
\eeq
plugging into \reef{Gappr}, and rescaling the integration variable, one can then fix $\gamma$ and $A$ \cite{Fitzpatrick:2012yx}:
\beq
\label{gammaA}
\gamma=4\Delta_\phi,\qquad A =\int_0^\infty dt\, t^{\gamma-1} K_0(2 t) =\Gamma(\gamma/2)^2/4\,.
\eeq

Of course Eq.~\reef{ptilde0} cannot be true literally, since we know that $p(\Delta)$ may contain a delta-function component. Instead, this is supposed to be true `on average'. A mathematically precise formulation is that the integrals of the two sides of \reef{ptilde0} should have the same asymptotics:
\beq
\label{ptilde}
Q(Y)=\int_0^Y d\Delta\,C(\Delta) p(\Delta)\sim (A\gamma)^{-1} Y^\gamma \qquad(Y \to\infty).
\eeq
This assumption leads to the same values of $\gamma$ and $A$ as the stronger assumption \reef{ptilde0}. To see this, one rewrites \reef{Gappr} in terms of $Q(Y)$ via integration by parts. 

As a simple consistency check, notice that the computation producing \reef{gammaA} is dominated by $x$, $\Delta$ such that $x\ll 1$, $\Delta\gg 1$ and $t=\sqrt{x} \Delta =O(1)$. In this region $x\Delta \ll 1$, as is needed for the validity of the approximation \reef{CBappr}. 

\begin{figure}
\centering
\includegraphics[width=0.4\textwidth]{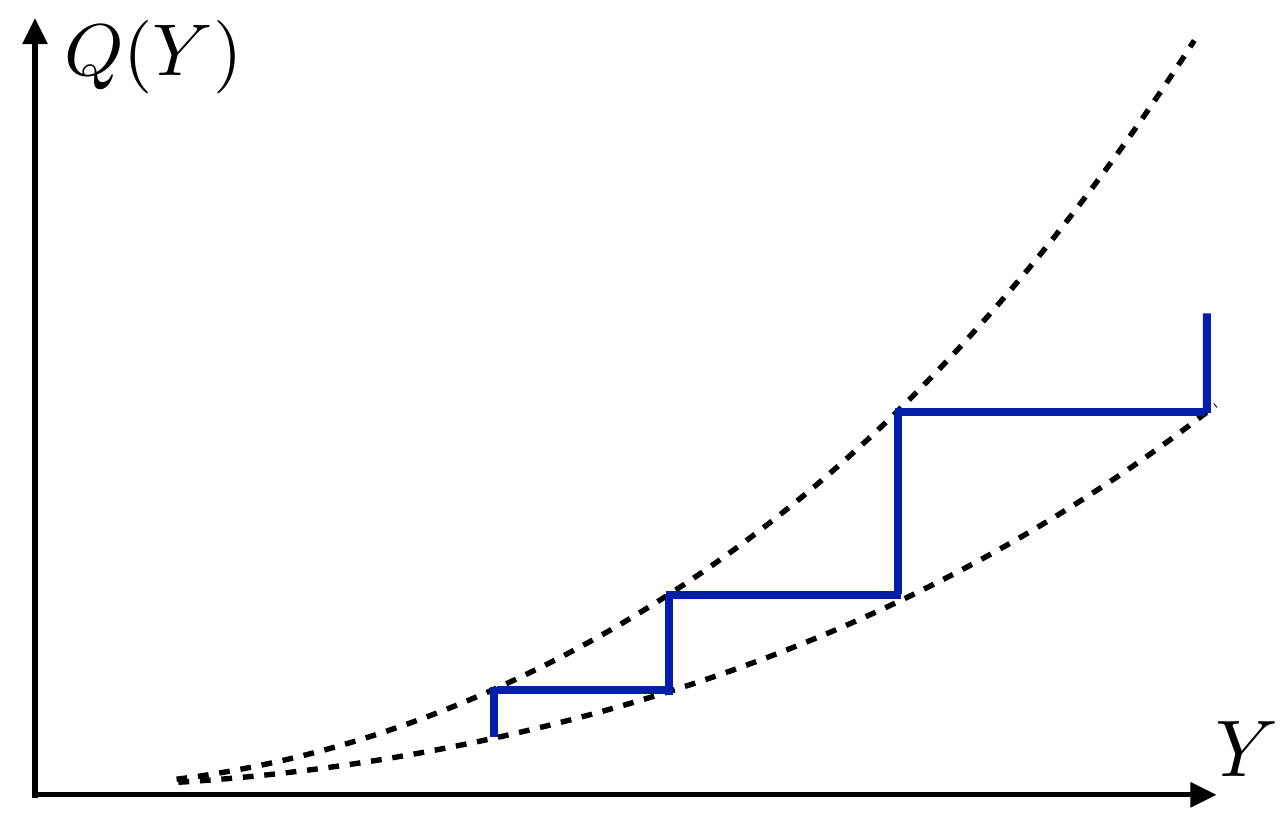}
\caption{Could $Q(Y)$ oscillate between two different powerlaw asymptotics? We will show that such a behavior is impossible.}
\label{could}
\end{figure}

What we have just reviewed is an appealing intuitive argument. Still, one may worry that Ref.~\cite{Fitzpatrick:2012yx} basically had to \emph{assume} the simple powerlaw asymptotics \reef{ptilde0}, or its mathematically precise form \reef{ptilde}. While this assumption is the simplest one consistent with powerlaw asymptotics in $x$, is it the only one? A priori, one can envision different kinds of behavior. For example, couldn't $Q(Y)$ oscillate between two different powerlaws as in figure~\ref{could}? 

The main purpose of our paper will be to explain that asymptotics \reef{ptilde}, with $\gamma$ and $A$ as in \reef{gammaA}, can in fact be obtained from \reef{CBdec} and \reef{zto1} without any further assumptions. The result derived in \cite{Fitzpatrick:2012yx} is thus true. In particular, the behavior shown in Fig.~\ref{could} is impossible.

The proof has the following structure. In the next section we explain why one can replace conformal blocks by their Bessel function approximations, as in Eq.~\reef{Gappr}. We then proceed to the crux of the problem, which is how \reef{Gappr} implies the asymptotics for the integrated spectral density. This will be shown using results from a branch of mathematics known as ``tauberian theory'' \cite{Korevaar}.\footnote{This theory takes its origin from a theorem about power series proved by an Austrian mathematician Alfred Tauber in 1897.} One well-known result of this kind is the Hardy-Littlewood tauberian theorem, which was already used in the CFT context in \cite{Pappadopulo:2012jk}. Here we need another tauberian theorem, for which we will give a self-contained explanation.

\section{From conformal blocks to the Bessel function}
\label{secBessel}

In this section we will supply a rigorous argument that \reef{CBdec} and \reef{zto1} imply \reef{Gappr}. 
The conformal block asymptotics involves two limits, large $\D$ and small $x$. In the rigorous argument it's convenient to separate them. First we simplify things by taking advantage of large $\D$, and then of small $x$.

\subsection{Large $\D$}
\label{sec:largeD}

We will need an asymptotics of conformal blocks in the large $\Delta$ limit, which is a slight refinement of the asymptotics used in \cite{Fitzpatrick:2012yx,Komargodski:2012ek}. Let us start from the integral 
representation\footnote{To get to the second line, change variables to $u=t/(1-t)$, rescale $u\to u/\sqrt{x}$, and use the invariance under $u\to u^{-1}$ to restrict integration to $u\ge 1$.}
\begin{align}
{}_2F_{1}(\D,\D,2\D,1-x)&= 
\frac{\Gamma(2\Delta)}{[\Gamma(\D)]^2} \int_0^1 dt\,\frac{t^{\D-1}}{1-t}
\left(1+ \frac{tx}{1-t}\right)^{-\D}\,\nn\\
&=\frac{\Gamma(2\Delta)}{[\Gamma(\D)]^2} \frac{2 I}{(1+\sqrt{x})^{2\D}},
\end{align}
where  
\beq
\label{I}
I = \int_1^\infty \frac{du}{u}(1+\kappa y)^{-\D},\qquad y=u^{-1}+u-2\,,\qquad
\kappa=\frac{\sqrt{x}}{(1+\sqrt{x})^2}.
\eeq

The prefactor has asymptotics
\beq
\frac{\Gamma(2\Delta)}{[\Gamma(\D)]^2} = \half C(\D)\times [1+O(1/\Delta)]\qquad(\D\to \infty).
\eeq
On the other hand, as shown in appendix \ref{appIJ}, the integral $I$ can be approximated for large $\Delta$ by replacing $(1+\kappa y)^{-\D}$ with $e^{-\kappa y \Delta}$, which evaluates to a Bessel function:
\begin{gather}
I = J\times[1+O(1/\Delta^{a})]\qquad(0<a<1\text{ arbitrary})\,,\label{IJ}\\
J = \int_1^\infty \frac{du}{u}e^{-\kappa y \D}= e^{2\kappa\D} K_0(2\kappa\D).
\end{gather}
The basic idea of the proof is to split the integral $I$ into two parts, at small and large $y$. For small $y$ we can safely replace $1+\kappa y$ by $e^{\kappa y}$, while the contribution from large $y$ is subleading.

We thus obtain the following large $\D$ asymptotics of conformal blocks:
\beq
\label{largeDeltaAs}
\frac{G_\D(1-x)}{C(\D)  f_\Delta(x)} = 1+O(1/\Delta^{a}) \qquad(\D\to \infty),
\eeq
uniformly in $x\in(0,1)$, where
\beq
\label{fD}
f_\D(x) = \left ( \frac{1-\sqrt{x}}{1+\sqrt{x}} \right ) ^\Delta 
e^{2\kappa\D} K_0(2\kappa\Delta).
\eeq

We now use this asymptotics to show that
\beq
\label{Fint}
\calF(x) = \int _{\Delta_0}^\infty d\D\,C(\D) p(\D)f_\D(x)\sim x^{-2\D_\phi}\qquad (x\to 0).
\eeq
Here $\Delta_0>0$ is any fixed number. Intuitively this follows from the fact that both this integral and \reef{CBdec} are dominated by large $\D$. Still, let us give a rigorous proof. 

We will show that $\lim_{x\to 0} \calG/\calF =1$. Pick a small $\eps>0$ and choose $\Delta_*$ such that the ratio in \reef{largeDeltaAs} stays close to 1 within $\pm\eps$ for $\Delta\ge \Delta_*$. Now split both integrals \reef{CBdec} and \reef{Fint} into two parts, $\calG_1$, $\calF_1$ below $\Delta_*$ and $\calG_2$, $\calF_2$ above $\Delta_*$. The parts $\calG_1$, $\calF_1$ can be bounded by a constant times $|\log x|$ for $x$ close to 0.\footnote{This follows from the fact that the individual conformal blocks satisfy such a bound with a uniform constant if $\Delta\in[0,\Delta_*]$. The same is true for $f_\Delta(x)$ if $\Delta\in[\D_0,\Delta_*]$.} On the other hand $\calG$, and hence $\calG_2$, grow in this limit as a powerlaw. It follows that the limit of $\calG/\calF$ is the same as the limit of $\calG_2/\calF_2$. The latter ratio stays close to 1 within $\pm\eps$ for any $x$, since the ratio of the integrands does so (here it's useful that the asymptotics \reef{largeDeltaAs} is uniform in $x$). So we conclude that the limit of $\calG/\calF$ is 1 within $\pm\eps$.\footnote{Strictly speaking we should have phrased this argument in terms of lim sup and lim inf. We will allow ourselves this imprecision several times in this paper.} Since $\eps$ was arbitrary, the limit is 1.

\subsection{Small $x$}
\label{sec:smallx}
Let us rewrite \reef{fD} as
\beq
\label{fDrewrite}
f_\D(x) = e^{-\phi \D}K_0(2\kappa\D),\qquad
\phi=-\log\left(\frac{1-\sqrt{x}}{1+\sqrt{x}}\right)-2\kappa,
\eeq
For small $x$ we have (notice that $\kappa\sim\sqrt{x}$)
\beq\label{phi}
\phi=4x+O(x^{3/2})=4\kappa^2+O(\kappa^3).
\eeq

We would like to argue that in the small $x$ limit the prefactor $e^{-\phi \D}$ in \reef{fDrewrite} can be dropped, so that \reef{Fint} implies
\beq
\label{intK}
\int _{\D_0}^\infty d\D\,C(\D) p(\D)K_0(2\kappa\D) \sim \kappa^{-4\D_\phi}\qquad(\kappa\to 0).
\eeq

In the intuitive reasoning of \cite{Fitzpatrick:2012yx,Komargodski:2012ek} reviewed in section \ref{formulation}, one first drops the prefactor, then runs the rest of the argument, and then comes back to check whether dropping was justified. This a posteriori consistency checks works out fine: the relevant values of $\Delta$ are $O(1/\sqrt{x})$, for which $\phi \D=O(\kappa)$, and so the prefactor is indeed close to one. This argument can be made rigorous using an idea outlined in \cite{Fitzpatrick:2014vua} around Eq.~(F.29).

Here we would like to offer a slightly different rigorous proof that the prefactor can be dropped, which is completely independent from the rest of the argument. The idea is straightforward---we compare the integrands of \reef{Fint} and \reef{intK}. One direction is simple. From \reef{phi} we know that $\phi>0$ for small $\kappa$, and so
\beq
f_\D(x) \le K_0(2\kappa\D).
\eeq
So the integrand of \reef{intK} is always larger than that of \reef{Fint}. So the asymptotics of \reef{intK} \emph{cannot be smaller} than what is shown in the r.h.s.

For the other direction, we would like to prove that an opposite inequality between the integrands is valid as long as we include a constant arbitrarily close to 1. There is also a useful freedom to somewhat rescale the argument of the integrand. There are many ways to implement this strategy. One possibility is as follows. Let us pick a small $\eps>0$. We claim that for sufficiently small $\kappa\le \kappa_*(\eps)$ there is an inequality
\beq
f_\D(x) \ge \frac{1}{1+C_1 \eps} K_0\bigl(2\kappa\D(1+C_2 \eps^2)\bigr)\qquad(\Delta\text{ arbitrary})
\eeq
with some $C_1,C_2>0$. If so, the asymptotics of \reef{intK} \emph{cannot be larger} than what is shown in the r.h.s., up to a factor of $1+O(\eps)$. Since $\eps$ is arbitrary, the asymptotics has to be exactly the one given in \reef{intK}.

To prove this last inequality, we will pick $\kappa_*=\eps^2$. Then for $\kappa\le \kappa_*$ we have
\beq
\label{ephi}
e^{-\phi \Delta} = e^{-O(\kappa)\kappa\Delta } = e^{-O(\eps^2) \kappa\Delta}.
\eeq
We consider two cases:

(a) $\kappa\Delta\le 1/\eps$. Then \reef{ephi} is $\ge 1/(1+C_1 \eps)$ and we are done. Notice that $K_0$ is monotonic so the increase of its argument only makes the inequality stronger.

(b) $\kappa\Delta\ge 1/\eps$. In this case we can use the asymptotics for $K_0$:
\beq
K_0(z)= (1+O(z^{-1}))\sqrt{\pi/(2 z)}e^{-z}\qquad(z\to\infty).
\eeq
So
\begin{align}
e^{-\phi \Delta} K_0(2\kappa\D) &= (1+O(\eps))\sqrt{\frac{\pi}{2 (2\kappa\D)}}e^{-[2+O(\eps^2)]\kappa\D} \\
&\ge (1+O(\eps)) K_0\bigl(2\kappa\D(1+C_2 \eps^2)\bigr)\,,
\end{align}
and we are done.

\section{Reduction to the tauberian theorem}
\label{red}

We would like to show that \reef{intK} implies \reef{ptilde}. Let us denote:
\beq
\label{notation0}
Y=\kappa^{-1},\qquad q(\Delta)=\begin{cases} 0,&\Delta\le \Delta_0\\
C(\Delta)p(\Delta),& \Delta>\Delta_0\end{cases}
\eeq
and introduce two ``weight functions":
\beq
\label{notation}
w_1(t)= K_0(2t),\qquad w_2(t) = \Theta(0\le t\le1),
\eeq
where $\Theta$ is the indicator function of the shown interval.

Then \reef{intK} can be rewritten as
\beq
\label{w1}
\int_0^\infty d\Delta\,q(\Delta) w_1(\Delta/Y) \sim Y^\gamma \qquad (Y\to \infty),
\eeq
while \reef{ptilde} with $A$ as in \reef{gammaA} takes the form:
\beq
\label{w2}
\int_0^\infty d\Delta\,q(\Delta)  w_2(\Delta/Y) \sim (I_2/I_1) Y^\gamma\qquad (Y\to \infty),
\eeq
where 
\beq
I_i = \int_0^\infty dt\,t^{\gamma-1}w_i(t).
\eeq

The integrals in the l.h.s.~of \reef{w1} and \reef{w2} express scaled weighted averages of the spectral density $q(\D)$. We need thus to show that we can replace one weight function with another by preserving the asymptotic behavior of the averages (after an appropriate rescaling). That this can be done, under certain conditions on the weight functions, is known in mathematics as a ``tauberian theorem".\footnote{The crux of this result is in the exact prefactor shown in \reef{w2}. If one is only interested in knowing that the l.h.s. of \reef{w2} is asymptotically bounded above and below by some constant times $Y^\gamma$, then an elementary proof can be given, see appendix \ref{sec:simple}.} 

Before we proceed to the proof, let us transform the statement of the theorem to a simpler form. First of all it will be convenient to transfer the dependence on $\gamma$ from the growth exponent in $Y^\gamma$ to the weight functions. To do this we introduce new spectral density and the weight functions:
\beq
q^{\rm new}(\Delta) = q(\Delta)/\Delta^{\gamma-1},\qquad w_i^{\rm new}(t)= t^{\gamma-1} w_i(t).
\eeq
It's easy to see that the tauberian theorem in terms of the new quantities takes the same form as before but with the growth exponent $\gamma=1$. We will also normalize the weight functions so that they have integral one. This is the theorem we will be proving:


\vspace{0.03in}

\noindent{\bf Tauberian theorem:} {\it Let $q(\Delta)$ be a non-negative spectral density and $w_1(t)$, $w_2(t)$ be two functions with unit integrals:
\beq 
\int _0^\infty dt\,w_i(t) =1.
\eeq
Suppose that 
\beq
\label{4.8}
Y^{-1}\int_0^\infty d\Delta\,q(\Delta) w_1(\Delta/Y) \sim 1 \qquad (Y\to \infty).
\eeq
Then, under certain extra conditions on $w_i(t)$ which will be made clear below,
\beq
\label{4.9}
Y^{-1}\int_0^\infty d\Delta\,q(\Delta)  w_2(\Delta/Y) \sim 1\qquad (Y\to \infty).
\eeq
}

\noindent For applications to the conformal bootstrap, we will need this theorem for
\beq
\label{willneed}
w_1(t) = \frac{4}{\Gamma(\gamma/2)^2} t^{\gamma-1} K_0(2t),\qquad w_2(t) = \gamma t^{\gamma-1} \Theta(0\le t\le1)\,.
\eeq

\section{Proof of the tauberian theorem}
\label{proof}

Tauberian theory is a rich branch of mathematics, see \cite{Korevaar} for a review. For a non-expert it may be hard to locate the needed result and to understand its proof. In fact the general theorem we need goes back to Wiener \cite{Wiener} (see e.g.~\cite{Korevaar}, Chapter II, Theorem 15.2). We found the exposition of Bochner \cite{Bochner,Bochner-collected} very clear (once we translated it from German). Our proof is a simplification of Bochner's (possible since we don't prove a general result but only what is needed for CFT applications). Of course we could have just cited mathematics literature and be done, but we believe that there is added value in seeing how things work. Since tauberian theorems are destined to continue to play a role in the conformal bootstrap, our simplified self-contained exposition will hopefully be useful.

\begin{figure}
\centering
\includegraphics[width=0.5\textwidth]{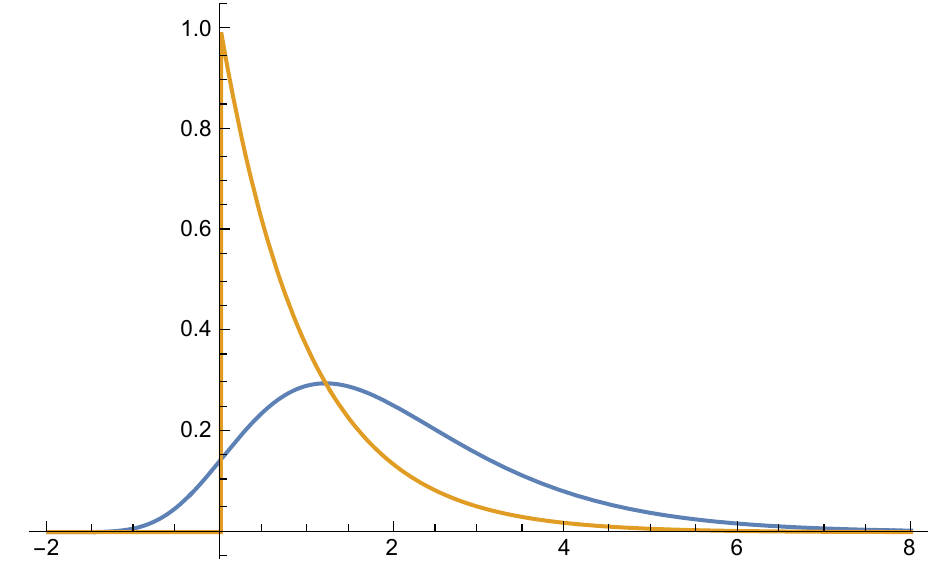}
\caption{The weight functions \reef{W1W2} (for $\gamma=1$).}
\label{W}
\end{figure}

The proof will rely on Fourier analysis. The integrals in the formulation of the theorem can be rewritten as ($w$ stands for any of the two weights)
\beq
\label{rew1}
Y^{-1}\int_0^\infty d\Delta\,q(\Delta) w (\Delta/Y) = \int_0^\infty \frac{d\Delta}{\D} \,q(\Delta) \tilde w (Y/\D),
\eeq
where
\beq
\tilde w(t)= t^{-1} w(t^{-1}).
\eeq 
In this form it is a multiplicative convolution of $q$ and $\tilde w$. It will be convenient to go to the usual additive convolution via change of variables
\beq
\Delta = e^x,\qquad Y=e^y,\qquad \rho(x) = q(e^x),\qquad W(x)= \tilde w(e^x) =e^{-x} w(e^{-x}).
\eeq
Notice that 
\beq
\int_{-\infty}^\infty dx\, W(x)=\int_0^\infty dt\,w(t) =1\,.
\eeq
In terms of the new variables \reef{rew1} becomes
\beq
\int_{-\infty}^\infty dx \,\rho (x) W (y-x) = (\rho*W)(y).
\eeq
So we achieved yet another reformulation of the Tauberian theorem. We are given, on the whole real line, a non-negative spectral density $\rho$ and two normalized weight functions $W_1$ and $W_2$. We know that
\beq
\label{W1}
\tag{$W_1$}
(\rho*W_1)(y)\to 1\qquad(y\to\infty) 
\eeq
and we need to show that
\beq
\label{W2}
\tag{$W_2$}
(\rho*W_2)(y)\to 1\qquad(y\to\infty).
\eeq
For applications to the conformal bootstrap, we will need this result for (see Fig.~\ref{W})
\beq
\label{W1W2}
W_1(x) = \frac{4}{\Gamma(\gamma/2)^2} e^{-\gamma x} K_0(2e^{-x}),\qquad W_2(x) = \gamma e^{-\gamma x} \Theta(x\ge 0)\,.
\eeq
We will actually explain the proof of the tauberian theorem only for these two functions. However, the given arguments will be sufficiently general, so that the reader will be able to adapt them to other functions of interest. See 
sections \ref{HL}, \ref{unequal}, \ref{largeN} for examples of such adaptations.

\begin{figure}
\centering
\includegraphics[width=0.5\textwidth]{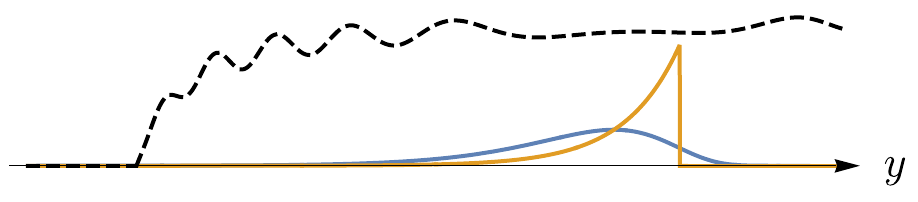}
\caption{An intuitive illustration, see the text. Notice that $\rho(x)$ vanishes for $x<\log \Delta_0$.}
\label{clear}
\end{figure}

Fig.~\ref{clear} provides an intuitive reason for the validity of this result. The oscillating dashed curve represents a non-negative spectral density (which in general does not have to be continuous and may contain a delta-function component). The averages of this spectral density with $W_1(y-x)$ go to a constant as $y\to\infty$. Intuitively this implies that the spectral density itself, roughly, goes to a constant. Then the averages with another weight function $W_2(y-x)$ should also go to a constant. 

\subsection{Proof that \reef{W2} implies \reef{W1}}
\label{secW2W1}

While our main goal is to prove that \reef{W1} $\Rightarrow$ \reef{W2}, we will start here by proving that \reef{W2} $\Rightarrow$ \reef{W1}. The key idea is to represent $W_1$ as a convolution of $W_2$ and some other function:
\beq
W_1 = W_2*R 
\label{R}
\eeq
It's easy to check that this equation is satisfied for (see note \ref{noteR} for how to find this $R$).
\beq
\label{Rexpl}
R(x) = W_1(x) + \gamma^{-1} W_1'(x).
\eeq
Notice that $R$ has integral one. Now, by the usual properties of convolution, we have
\beq
\label{eq:conv}
\rho * W_1 = (\rho* W_2)*R.
\eeq  
Interchanges of the order of integration are easy to justify.

 Denote $u = \rho*W_2$. Notice that while $\rho$ is a spectral density which may have a delta-function component, $u$ is a piecewise-continuous function. Also, this function is uniformly bounded, i.e.~there exists a constant $C$ so that
\beq
0\le u(x) <C\text{ for all }x\,.
\eeq
Indeed, for large $x$ it is uniformly bounded because it approaches $A$. On the other hand, for $x$ below some $x_0$, we can bound $u(x)$ by a constant times $u(x_0)$. {To show this, we use the inequality:
\beq
\label{eq:const}
W_2(x-t)\le const.\, W_2(x_0-t)\qquad(t>t_0)\,,
\eeq
where $t_0$ is such that $\rho(t)=0$ for $t\le t_0$.}

Given \reef{eq:conv}, we have to prove that if $u\to A$ as $y\to\infty$, then $u*R\to A$ in the same limit. We are interested in $A=1$, but passing to $u^{\rm new}=u-A$, $\rho^{\rm new}=\rho-A$ we are reduced to the case $A=0$.
The new $u$ is still uniformly bounded, although not necessarily non-negative.

The rest is easy. Pick any $\eps>0$. Find an $x_0$ such that $|u(x)|<\eps$ for $x>x_0$. We can split the convolution $u*R(y)$ as follows:
\beq
u*R(y) = \int_{-\infty}^{\infty} dx\, u(x)R(y-x)=\int_{x<x_0}+\int_{x>x_0}.
\eeq 
In the first integral $|u|<C$, so it is bounded by 
\beq
C \int_{y-x_0}^{+\infty} dt\, |R(t)|
\eeq 
and goes to zero as $y\to\infty$. In the second integral $|u|<\eps$ so it's bounded by
\beq
\eps \int_{-\infty}^{+\infty} dt\, |R(t)| = O(\eps).
\eeq
We have just shown that $\lim_{y\to\infty} u*R(y)=O(\eps)$ for any $\eps$, so this limit is zero. QED.

The reader will notice that the non-negativity of $\rho$ was not essential here. It would be sufficient to assume that the integral defining the convolution \reef{W2} is absolutely convergent for any $y$. This is unlike the opposite implication for which $\rho\ge0$ will be crucial. 

The used properties of $W_1$ and $W_2$ were: the existence of the convolution representation \reef{R},
{and the inequality \reef{eq:const} needed to argue for the boundedness of $u$. }

\subsection{Proof that \reef{W1} implies \reef{W2}}

If we try to apply the proof from the previous section to show the opposite implication \reef{W1} $\Rightarrow$ \reef{W2} we encounter a difficulty: in contrast to \reef{R}, there is no function $S$ such that
\beq
\label{W2W1S}
W_2 = W_1 * S.
\eeq
This is obvious already from the fact that $W_2$ has a discontinuity, while $W_1$ is smooth. Whatever an integrable function $S$, the convolution $W_1*S$ will be necessarily a continuous function, and so cannot equal $W_2$.

We can reach the same conclusion using the Fourier transform. In Fourier space Eq.~\reef{W2W1S} becomes
\beq
\hat W_2(p) = \hat W_1(p) \hat S(p).
\eeq
The Fourier transforms are given by:\footnote{Changing variables back to $t=e^{-x}$ these become Mellin transforms of $w_1(t)$, $w_2(t)$ given in \reef{willneed}.}
\begin{gather}
\label{MW1}
\hat W_1(p)=\Gamma\left(\half(\gamma+ip)\right)^2\left/\Gamma\left(\half \gamma\right)^2\right.,\\
\hat W_2(p) = \frac{\gamma}{\gamma+ip}.
\end{gather}
The $\hat W_1(p)$ decays exponentially fast at large $p$:
\beq
|\hat W_1(p)|\sim const. |p|^{2\gamma-1} e^{-\pi|p|}\qquad(p\to\infty),
\eeq
while $\hat W_2(p)$ decays only as $p^{-1}$ (which is related to the fact that $W_2(x)$ is discontinuous). So if the function $S$ existed, its Fourier transform would have to grow exponentially at infinity, a contradiction.\footnote{\label{noteR}On the other hand the function $R$ solving \reef{R} must have the Fourier transform
$\hat R(p) = \hat W_1(p)/\hat W_2(p) = \hat W_1(p)(1+ip/\gamma).$
Inverting this, we get \reef{Rexpl}.}

We will now explain how one can work around this difficulty. 
The key idea is as follows. Fix a small $\eps>0$. We claim that we can find a pair of functions $W_2^{\pm}$ such that
\begin{gather} 
W_2^-(x)\le W_2(x) \le W_2^+(x)\qquad\text {for any }x,
\label{squeeze}\\
\left|\int dx\,W_2^\pm (x) - 1\right| <\eps,
\label{squeeze1}
\end{gather}
(recall that $W_2$ has integral 1) while the Fourier transforms 
\beq
\hat W_2^\pm(p) \qquad\text{decay as $O(e^{-const.p^2})$ at large $p$}.
\eeq

From the existence of these functions, let us see how we can finish the proof.
First of all we can use the argument of section \ref{secW2W1} to show that \reef{W1} implies ($W_2^\pm$), by which we mean:
\beq
\rho* W_2^\pm(y)\to \int dx\,W_2^\pm(x)\qquad(y\to\infty).
\eeq
We define the function $R$ by solving the convolution equation in Fourier space:
\beq
\label{Rpm}
\hat R(p) = \hat W_2^\pm(p)/\hat W_1(p).
\eeq 
This solution makes sense because the Fourier transform of $W_2^\pm$ decays at large $p$ faster than that of $W_1$, and because $\hat W_1(p)$ does not vanish for any $p$.\footnote{Also notice that condition \reef{eq:const} used in section \ref{secW2W1} is satisfied for $W_1$ in place of $W_2$.}

Second, we observe that, for any $y$
\beq
\label{rhocomp}
\rho* W_2^-(y) \le \rho* W_2(y) \le \rho* W_2^+(y).
\eeq
It is in this step that we use the fact that the spectral density $\rho$ is non-negative.

Since the functions in the l.h.s.~and in the r.h.s.~of the last inequality tend to 1 within $\pm\eps$ as $y\to\infty$,
and since $\eps$ is arbitrary, we conclude that the function in the middle must tend to 1. 

\subsection{Construction of functions $W_2^{\pm}$}

\begin{figure}
\centering
\includegraphics[width=0.5\textwidth]{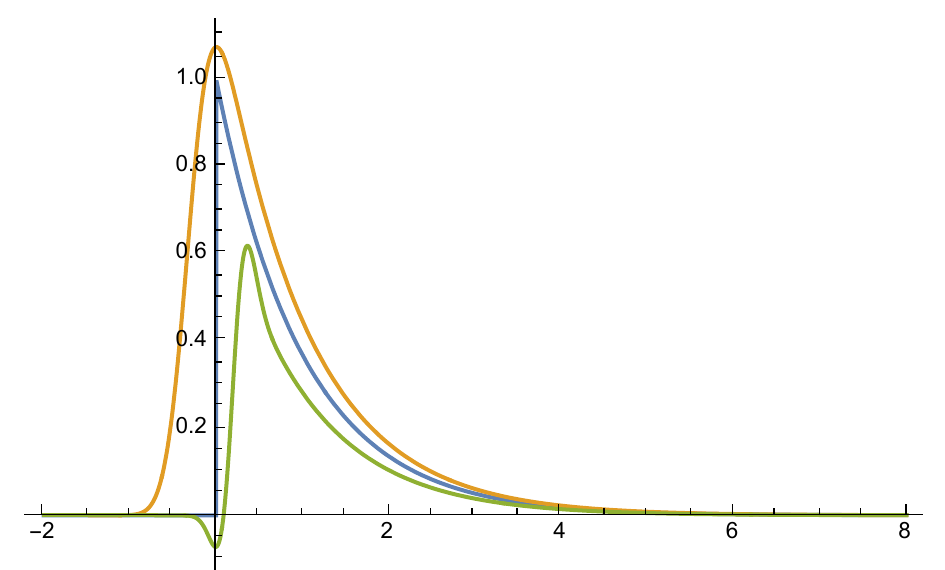}
\caption{Approximating the function $W_2$ from above and below by functions $W_2^\pm$.}
\label{Wpm}
\end{figure}

The functions $W_2^\pm$ should look like in Fig.~\ref{Wpm}, i.e.~they should be smooth functions closely approximating $W_2$ from above and below. We construct these functions by taking some approximations of $W_2$ from above and below, and then smoothing these approximations by convolution. One has to work a bit to make sure that after the convolution the pointwise comparison inequalities \reef{squeeze} are still satisfied.

Consider auxiliary functions $\phi^\pm$ with the following properties:
\begin{gather}
\int_{|x|<\eps} dx\,\phi^\pm (x)=1,
\label{intphi1}\\
\phi^+(x)>0,\phi^-(x)<0\qquad \text{ for }|x|>\eps,
\label{tailphi}\\
\int_{-\infty}^\infty dx\,\phi^\pm(x)=1\pm O(\eps),
\end{gather}
and, finally, the Fourier transform of $\phi^\pm$ decays as $O(e^{-const.p^2})$. 
Such functions can be found within a family of functions $const. e^{-(x/a)^2}$ for $\phi^+$ and
$const. e^{-(x/a)^2}(1-(x/b)^2)$ for $\phi^-$.

The functions $W_2^\pm$ can then be defined as
\begin{gather}
W_2^\pm(x) = \phi^\pm * V^\pm,\\
V^+(x)=\max_{y\in[x-\eps,x+\eps]} W_2(y),\\
V^-(x)=\min_{y\in[x-\eps,x+\eps]} W_2(y).
\end{gather}
The property \reef{squeeze} follows from \reef{intphi1} and \reef{tailphi}, and from the fact that $W_2$ is everywhere non-negative. At the same time we have
\beq
\int W_2^{\pm} = \left(\int \phi^\pm\right)\left(\int V^\pm\right)=1\pm O(\eps).
\eeq

\section{Comments and extensions}

\subsection{Nonvanishing Fourier transform}
\label{nonvanish}

As we have seen, the main idea in the proof of the tauberian theorem is to represent some weight functions as convolutions of other weight functions. These representations are found with the use of Fourier transform.
In our proof, a key step was that that Eq.~\reef{Rpm} defines a Fourier transform of a function. For this it was important that 
\begin{enumerate}
\item
$\hat W_1(p)$ nowhere vanishes.
\item
$\hat W_1(p) e^{\eps |p|^2}\to\infty$ for any $\eps>0$.
\end{enumerate}
It turns out that condition 2 is not important and can be removed at the cost of complicating the proof \cite{Bochner,Bochner-collected}.
On the other hand condition 1 is crucial and in its absence the tauberian theorem cannot hold. Indeed, suppose that we have a normalized weight function $W_1$ such that its Fourier transform has a zero at some $p_0$: $\hat W_1(p_0)=0$. We can consider a non-negative spectral density
\beq
\rho(x)= 1+\cos(p_0 x).
\eeq
The condition \reef{W1} is then satisfied. On the other hand, the r.h.s. of \reef{W2} behaves as
\beq
\rho*W_2(y)= 1+ \half [e^{ip_0 y} \hat W_2(p_0)+ \text{c.c.}]
\eeq
and does not go to a constant.

\subsection{Hardy-Littlewood tauberian theorem}
\label{HL}

Another important tauberian theorem which occurs in CFT applications is the Hardy-Littlewood theorem, which was already invoked in \cite{Pappadopulo:2012jk}. Let us recall how it arises. We can take the conformal block decomposition of the 4pt function, Eq.~\reef{CBdec}, and split it into simple powers:
\beq
\label{Gr}
\calG(z) = \int_0^\infty d\Delta\, r(\Delta)\, z^\Delta.
\eeq
To do this we just take each conformal block and expand it into powers. The coefficients of this expansion being positive, we conclude that $r(\Delta)$ is still a non-negative spectral density. The interpretation of this operation is that $\Delta$ in \reef{Gr} runs over primaries and descendants, while in \reef{CBdec} it was numbering the primaries only.

Now let's forget about the origin of $r(\Delta)$ apart from it being non-negative, and just take a general power series \reef{Gr} which satisfies the asymptotics \reef{zto1}. The Hardy-Littlewood theorem states that, under these conditions, the spectral density satisfies the integrated asymptotics:
\begin{gather}
\int_0^Y d\Delta\,r(\Delta) \sim (A\gamma)^{-1} Y^\gamma,\\
\gamma = 2\Delta_\phi,\qquad A=\int_0^\infty dt\,t^{\gamma-1} e^{-t} =\Gamma(\gamma).
\end{gather}
To see this, we introduce the variable $t=|\log z|$ and rewrite the conditions of the theorem as
\beq
\int_0^\infty d\Delta\, r(\Delta)\, e^{-\Delta t}\sim t^{-2\Delta_\phi}\qquad(t\to0).
\eeq
The analogy with Eqs.~\reef{Gappr}, \reef{ptilde} and \reef{gammaA} should now be clear.

For a rigorous proof, we should run the argument in section \ref{proof}. The difference is that now we have a different weight function:
\beq
w_1(t) = \frac{1}{\Gamma(\gamma)} t^{\gamma-1} e^{-t}
\eeq
and the corresponding
\beq
W_1(x) = \frac{1}{\Gamma(\gamma)} e^{-\gamma x} e^{-e^{-x}}.
\eeq
The Fourier transform is given by
\beq
\label{MW1HL}
\hat W_1(p) = \frac{\Gamma(\gamma+ip)}{\Gamma(\gamma)}.
\eeq
This Fourier transform satisfies both conditions 1,2 emphasized in section \ref{nonvanish}. So our argument is sufficient to prove the Hardy-Littlewood theorem.

In appendix \ref{Karamata} we discuss another proof of the Hardy-Littlewood theorem due to Karamata, and why it does not quite work to prove the conformal bootstrap tauberian theorem. 

\subsection{Positivity of the spectral density}

\begin{figure}
\centering
\includegraphics[width=0.4\textwidth]{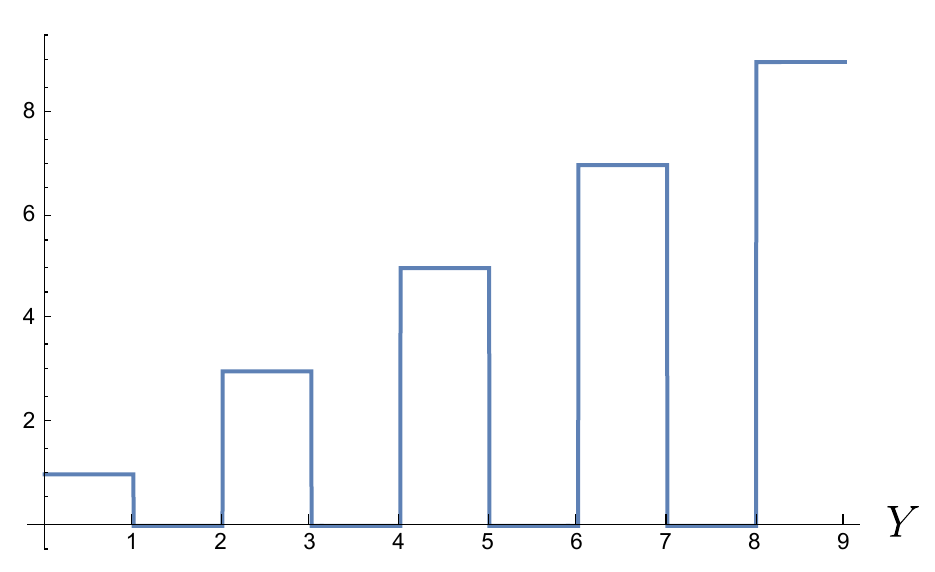}
\caption{The integrated spectral density $\int_0^Y d\Delta\,r(\D)$ corresponding to the series \reef{ex}.}
\label{density}
\end{figure}

Positivity of the spectral density is important for the tauberian theorems to hold. We will demonstrate this on the example of the Hardy-Littlewood theorem. Consider the function \cite{Titchmarsh}:
\beq
\label{ex}
\calG(z) = \frac{1}{(1-z)(1+z)^2} = 1-z+2 z^2-2 z^3+3 z^4-3 z^5+4 z^6-4 z^7+\ldots
\eeq
The integrated spectral density oscillates as shown in Fig.~\ref{density} as opposed to growing asymptotically linearly.

\subsection{Generalization to unequal external dimensions}
\label{unequal}

We would like to give a quick but rigorous discussion of the case when the external operators have unequal dimensions.\footnote{See \cite{Fitzpatrick:2014vua} for the original discussion at an intuitive level.} 
The 4pt function in this case has the form
\beq
\langle \phi_1(0) \phi_2(z) \phi_2(1) \phi_1(\infty)\rangle = z^{-\D_1-\D_2} \calG(z)
\eeq
with the conformal block decomposition
\beq
\calG(z) = \int_0^\infty d\Delta\, p(\Delta)\, G^{(\delta)}_{\Delta}(z),\qquad G^{(\delta)}_\D(z)= z^\D{}_2F_{1}(\D-\delta,\D-\delta,2\D,z)\,,
\label{CBdecdelta}
\eeq
where $\delta=\Delta_1-\Delta_2$ \cite{DO3}.

We are considering a reflection-positive configuration so that the spectral density $p(\Delta)$ is non-negative. From the crossed channel we know that $\calG(1-x)\sim x^{-2\D_2}$ as $x\to 0$ and we would like to know what this implies for the asymptotic spectral density. We can assume without loss of generality that $\delta>0$, otherwise we apply an ${\rm SL}(2,\bR)$ transformation which interchanges $\phi_1$ and $\phi_2$. Mathematically, the cases $\delta>0$ and $\delta<0$ are equivalent because we have a hypergeometric identity
\beq
{}_2F_{1}(\D-\delta,\D-\delta,2\D,1-x) = x^{2\delta} {}_2F_{1}(\D+\delta,\D+\delta,2\D,1-x)\,.
\eeq
 
We need an approximation formula for the conformal blocks in the large $\Delta$ limit.
Going through the same steps as in section \ref{sec:largeD}, and using the same integration variables, the Euler integral representation for the hypergeometric function gives:
\beq
\label{Euler}
{}_2F_{1}(\D-\delta,\D-\delta,2\D,1-x)= 
\frac{\Gamma(2\Delta)}{\Gamma(\D-\delta) \Gamma(\D+\delta)} 
\frac{I_\delta}{(1+\sqrt{x})^{2\D-2\delta}}\,,
\eeq
where
\beq
\label{Idelta}
I_\delta=\int_0^\infty \frac{du}{u}\frac{1}{(1+u/\sqrt{x})^{2\delta} }(1+\kappa y)^{-\D+\delta}\,.
\eeq
The prefactor in \reef{Euler} has the same asymptotics $\half C(\D)$ as for $\delta=0$.
Rigorous analysis of the integral $I_\delta$ is simplified by using the following observation. Although non-obvious from this representation, it follows from the above hypergeometric identity that we have
 \beq
 \label{non-obv}
 \frac{x^{2\delta}}{(1+\sqrt{x})^{4\delta}}I_{-\delta} = I_{\delta}\,.
\eeq
Suppose now that we drop 1 in $(1+u/\sqrt{x})$ in \reef{Idelta}. Given that $\delta>0$, this gives an upper bound for $I_\delta$ and a lower bound for $I_{-\delta}$. Using \reef{non-obv}, we obtain a two-sided bound for $I_\delta$:
\beq
\frac{x^{\delta}}{(1+\sqrt{x})^{4\delta}} \int_0^\infty \frac{du}{u^{1-2\delta}}(1+\kappa y)^{-\D-\delta} \le  I_\delta \le  
{x^\delta} \int_0^\infty \frac{du}{u^{1+2\delta}}(1+\kappa y)^{-\D+\delta}\,.
 \eeq
Now by an argument similar to the one given in appendix \ref{appIJ} we can show that in the large $\Delta$ limit it's possible to replace $(1+\kappa y)$ by $e^{\kappa y}$ in the integrals on the right and on the left, committing a relative error at most $O(1/\Delta^{a})$, $0<a<1$.\footnote{This argument becomes a bit more tedious because in general for $\delta>0$ we cannot take advantage of the monotonicity of the integration measure in the proof of \reef{I2I1}, as in Eqs.~\reef{monot}. Nevertheless the statement remains true.} The resulting integrals then evaluate to another Bessel function, so that we obtain, up to relative error $O(1/\Delta^{a})$,
\beq
\frac{2x^{\delta}}{(1+\sqrt{x})^{4\delta}}  e^{2\kappa \D_+} K_{2\delta}(2\kappa \D_+) \le  I_\delta \le  2x^\delta  e^{2\kappa \D_-} K_{2\delta}(2\kappa \D_-)\,,
\eeq
where we introduced $\Delta_\pm=\Delta\pm\delta$. If we now estimate how much the r.h.s.~and the l.h.s.~vary due to this little shift of dimensions, we obtain:
\beq 
I_\delta = 2x^\delta  e^{2\kappa \D } K_{2\delta}(2\kappa \D)\times[1+O(\kappa)+O(1/\Delta^{a})]\,,
\eeq
and so finally:
\begin{gather}
 \label{largeDeltaAsdelta}
\frac{G^{(\delta)}_\D(1-x)}{C(\D)  f^{(\delta)}_\Delta(x)} = 1+O(\sqrt{x})+O(1/\Delta^{a}) \qquad(x\to0,\D\to \infty),\\
\label{fDdelta}
f^{(\delta)}_\D(x) = \left ( \frac{1-\sqrt{x}}{1+\sqrt{x}} \right ) ^\Delta 
e^{2\kappa\D} x^\delta  K_{2\delta}(2\kappa\Delta).
\end{gather}

Continuing to mimic the line of reasoning in section \ref{secBessel}, we now want to show that we can replace conformal blocks by $f^{(\delta)}_\D(x)$ in the asymptotics, i.e.~the analogue of \reef{Fint}:
\beq
\label{Fintdelta}
\calF(x) = \int _{\Delta_0}^\infty d\D\,C(\D) p(\D)f^{(\delta)}_\D(x)\sim x^{-2\D_2}\qquad (x\to 0).
\eeq
There is a minor difference that while \reef{largeDeltaAs} was true uniformly in $x$, our asymptotics \reef{largeDeltaAsdelta} contains an error term $O(\sqrt{x})$. However as we will see this will not be a problem.\footnote{On the other hand it's crucial that \reef{largeDeltaAsdelta} does not contain mixed error terms which decrease in $x$ but grow in $\Delta$, like e.g.~$O(x \Delta)$.} 

Here is the appropriate version of the last paragraph of section \ref{sec:largeD} (modifications in blue).
We will show that $\lim_{x\to 0} \calG/\calF =1$. Pick a small $\eps>0$ and choose $\Delta_*$ {\blue and $x_*$} such that the ratio in \reef{largeDeltaAsdelta} stays close to 1 within $\pm\eps$ for $\Delta\ge \Delta_*$ {\blue and $x\le x_*$}. Now split both integrals \reef{CBdecdelta} and \reef{Fintdelta} into two parts, $\calG_1$, $\calF_1$ below $\Delta_*$ and $\calG_2$, $\calF_2$ above $\Delta_*$. The parts $\calG_1$, $\calF_1$ can be bounded by {\blue a constant} for $x$ close to 0. On the other hand $\calG$, and hence $\calG_2$, grow in this limit as a powerlaw. It follows that the limit of $\calG/\calF$ is the same as the limit of $\calG_2/\calF_2$. The latter ratio stays close to 1 within $\pm\eps$ for {\blue $x\le x_*$}, since the ratio of the integrands does so. So we conclude that the limit of $\calG/\calF$ is 1 within $\pm\eps$.

Furthermore, arguing exactly like in section \ref{sec:smallx}, we show that $f^{(\delta)}_\D(x)$ can be replaced in \reef{Fintdelta} by $x^\delta  K_{2\delta}(2\kappa\Delta)$. Thus we finally obtain a condition of the tauberian type:
\beq
\label{eq:uneq-prob}
 \int _{\Delta_0}^\infty d\Delta\, C(\D)\, p(\D) K_{2\delta}(2\kappa \D)\sim x^{-2\D_2-\delta}= x^{-\D_1-\D_2}\sim \kappa^{-\gamma}\,,
\eeq
where $\gamma=2(\D_1+\D_2)$. As expected, this final equation is invariant under $\delta\to-\delta$.

We can now analyze the asymptotics of the spectral density using the methods of sections \ref{red} and \ref{proof}.
The weight function $w_1(t)$ in \reef{willneed} has to be replaced by
\beq
\label{torepl}
w_1(t)=t^{\gamma-1} K_{2\delta}(2t)
\eeq
(times a normalization factor). Notice that this weight function is integrable as $K_{2\delta}(2t)=O(t^{-2\delta})$ and $2\delta<\gamma$.

The resulting asymptotics for the spectral density will take the form:
\beq
\int_0^Y d\Delta\, C(\D) p(\D) \sim (A\gamma)^{-1}Y^\gamma,
\eeq
where 
\beq
A=\int_0^\infty dt\,t^{\gamma-1} K_{2\delta}(2t) = \frac{1}{4} \Gamma
   \left(\frac{\gamma }{2}-\delta
   \right) \Gamma \left(\frac{\gamma
   }{2}+\delta \right)\,.
\eeq

The Mellin transform of the weight function is given by:
\beq
\int_0^\infty dt\,t^{\gamma-1+i p } K_{2\delta}(2 t) = \frac14 \Gamma
   \left(\frac{\gamma+ip }{2}-\delta
   \right) \Gamma \left(\frac{\gamma+ip
   }{2}+\delta \right)\,.
\eeq
It is non-vanishing and has the same decay properties as the Mellin transform for the $\delta=0$ case. So our proof of the tauberian theorem goes through.

\subsection{Application to large $N$ gauge theories}

\label{largeN}
In this section we will change gears completely, and consider a confining 4d gauge theory in the limit of a large number of colors.\footnote{This additional application of the tauberian theory in physics was suggested to us by Sasha Zhiboedov.} In such a theory consider a local operator $\calO$, a scalar for simplicity. Its 2pt function in the Euclidean momentum space has a spectral representation \cite{Witten:1979kh} 
\beq
\langle \calO(p) \calO(-p)\rangle = \sum _n \frac{|c_n|^2}{p^2+m_n^2}\,.
\eeq
Here $m_n$ are the masses of resonances (infinitely narrow in the large $N$ limit), and $c_n$ are their couplings to $\calO$. The important point is that, in the strict $N\to\infty$ limit, this equation is supposed to apply at all momenta, including the asymptotically large momenta where the theory approaches the UV. At those momenta the 2pt function must scale as
\beq
\langle \calO(p) \calO(-p)\rangle \sim const. (p^2)^{\Delta-2}\,, \qquad (p^2\to\infty)\,,
\eeq
where $\Delta$ is the UV dimension of $\calO$, times a function which varies logarithmically with $p^2$ which is generated by the logarithmic running of the gauge coupling near the UV {and by the Fourier transform from the position space if $\Delta$ is an even integer}.\footnote{There may be extra factors which scale as even slower varying functions, like double logs etc. We will not write such factors but their presence will be compatible with the discussion below.}

While the above equations are usually written in momentum space, it will be more convenient for us to transform them to position space. This is because the series in momentum space may not converge without subtraction needed to eliminate the contact terms, while the series in position space should converge. Thus we obtain:
\beq
\langle \calO(0) \calO(x)\rangle = \sum _n |c_n|^2 \frac{1}{4\pi^2 x^2} w(m_n |x|) \sim \frac{const.}{(x^2)^\Delta}\,\qquad (x\to 0)\,,
\eeq
where
\beq
w(t)=t K_1(t)\,,
\eeq
and once again an extra logarithmic factor is left implicit.

Let us introduce the spectral density 
\beq
\rho(t)= \frac 1{4\pi^2}\sum_n |c_n|^2 \delta (t-m_n)
\eeq
and denote $Y=1/|x|$. Then we can rewrite the above equation as
\beq
\label{rewr}
\int_0^\infty dt\,\rho(t) w(t/Y)\sim const.  Y^{2\Delta-2} L(Y) \qquad (Y\to \infty)\,,
\eeq
where we now introduced explicitly the extra logarithmic factor $L(Y)$ . Notice that the exponent $2\Delta-2$ is positive in 4d by the unitarity bound. 

If there were no extra factor, $L(Y)\equiv 1$, then Eq.~\reef{rewr} would be of precisely the same form as \reef{w1}. The weight function is a partial case of \reef{torepl} analyzed in the previous section and satisfies the requirements needed for the application of the tauberian theorem. Thus we would obtain a rigorous result for the asymptotics of the integrated spectral density.

Turning now to the case with the extra factor, it's important that this factor satisfies the so called \emph{slow variation} condition
\beq 
\lim_{Y\to\infty} L(\lambda Y)/L(Y) = 1\qquad\text{for any $\lambda>0$}\,.
\eeq
For such factors there is a generalized tauberian theorem, Theorem 9.3 in \cite{Korevaar}, Chapter IV. Roughly, it says that the theorem in section \ref{red} remains true if we replace 1 by $L(Y)$ in the r.h.s.~of both \reef{4.8} and \reef{4.9}. So the conclusion is that the integrated spectral density in this case will have a powerlike asymptotics times $L(Y)$.

{The spectral density asymptotics in 4d large $N$ theories were recently discussed in \cite{Bochicchio:2013eda}, section 3. The spectral density asymptotics is recovered in that work by exhibiting a smooth spectral density which, upon doing the integral in $t$, gives asymptotics as $Y\to\infty$ which agrees with the r.h.s.~of Eq.~\reef{rewr} (for an appropriate $L(Y)$). An attempt is then made to prove that that this spectral density asymptotics is unique. Unfortunately, this part of the discussion in \cite{Bochicchio:2013eda} appears incorrect.\footnote{First, the use of Euler-Maclaurin formula in (3.10) presupposes that the sequence of pole residues can be analytically continued, which is not a given. Second, the whole attempt of justifying the uniqueness based on appealing to Fredholm alternative, Eq.~(3.13) and below, is unfounded, since the Fredholm alternative deals with full solutions of integral equations and not with asymptotic solutions. We thank Marco Bochicchio for having tried to convince us, unsuccessfully, in the correctness of his proof.} On the other hand, the above considerations based on the tauberian theorems provide an alternative and mathematically precise way to put the results of \cite{Bochicchio:2013eda} on solid grounds.}

See also \cite{Vladimirov:1978xx} for other applications of tauberian theorems in general quantum field theories.

\section{Discussion}

Conformal field theory is physically relevant. In addition, its equations are mathematically well defined. 
One advantage of having well-defined equations is that we can study them numerically. The modern developments in the conformal bootstrap program benefitted from the latter fact enormously, starting from \cite{Rattazzi:2008pe}. Many rigorous bounds on the parameter space of CFTs in various dimensions have since been obtained, perhaps the most impressive result being an accurate determinations of the low-lying CFT data for the 3d Ising model \cite{ElShowk:2012ht,El-Showk:2014dwa,Kos:2014bka,Simmons-Duffin:2015qma,Kos:2016ysd,Simmons-Duffin:2016wlq}. While these bounds and results have been obtained numerically, they have rigorous error bars, precisely because they follow from a well-defined set of equation. In particular, the theorem about OPE convergence proved in \cite{Pappadopulo:2012jk} provides solid basis for this numerical analysis (see also the recent comments in \cite{Rychkov:2017tpc}).

One point of view on the numerical results is that they are but the first glimpse of a dazzling future theory which will provide an analytic solution of the bootstrap equations. Another, more modest point of view, is that the bootstrap equations are too hard to solve analytically in most cases of physical interest. Still, we can make progress by improving our numerical techniques. One way to improve the existing numerical algorithms is to combine a numerical approach to the low-lying 
CFT data with analytic control over the high-dimension operators. This analytic control comes in particular from the lightcone bootstrap, which is the $d>1$ counterpart of the problem considered in this paper, reviewed in appendix \ref{d>1}. Looking for such a hybrid approach is a worthy goal, and some steps in this direction have already been taken recently in \cite{Simmons-Duffin:2016wlq,Caron-Huot:2017vep}. 

It is to be hoped that this future hybrid approach, when it is found, will preserve the nice feature of the current numerical approach in that the obtained bounds on the CFT parameter space will still be completely rigorous. If this is to happen, we are obliged to find a rigorous understanding of the high-dimension spectrum. In this paper we provided such an understanding for one toy problem in the context of 1d CFTs. 

It should be pointed out that there is still a huge gap between what we have shown and what is expected to be true, even in 1d CFTs. Here we established just the leading asymptotics of the integrated spectral density. But instead, it seems reasonable to expect that the spectrum of exchanged operators should become asymptotically equally spaced, approaching generalized free theory spectra discussed in appendix \ref{GFF} (or perhaps with a finite number of trajectories of this type). The distribution of these operators should respect the leading as well as subleading asymptotics corresponding to subleading terms in the OPE in the crossed channel. We expect that to prove such a result it will be crucial to use analytic structure of the 4pt function for complex $z$, similarly to what was done recently in \cite{Caron-Huot:2017vep}. Analyticity should lead to more powerful conclusions than what we have achieved here via real analysis methods (notice in particular that asymptotics \reef{zto1} was used by us only for $x\to0$ along the real axis).

\section*{Acknowledgements}

We are grateful to Marco Bochicchio, Liam Fitzpatrick, Jared Kaplan and David Simmons-Duffin for clarifications concerning their work, and to Sasha Zhiboedov for suggesting the application to large $N$ theories and for the comments on the draft. JQ is grateful to the CERN Theoretical Physics Department for hospitality. SR is supported by the National Centre of Competence in Research SwissMAP funded by the Swiss National Science Foundation, by the Simons Foundation grant 488655 (Simons collaboration on the Non-perturbative bootstrap), and by Mitsubishi Heavy Industries as an ENS-MHI Chair holder.

\appendix
\section{CFT in $d=1$}
\label{CFT1}

In this paper by unitary CFT in $d=1$ dimension we mean a system of correlation functions of local operators on the real line 
\beq
\langle \calO_1(x_1)\ldots \calO_n(x_n) \rangle\qquad (x_1<x_2<\ldots<x_n) 
\eeq
satisfying the following axioms which are adaptations of the axioms of global conformal invariance in $d>1$ (see e.g.~\cite{Rychkov:2016iqz}).

{\bf Axiom 1.} There is a privileged class of operators called primaries. Their correlation functions remain invariant under the ${\rm SL}(2,\bR)$ group of fractional linear transformations
\beq
x\to x'=f(x) = \frac{ax+b}{cx+d}
\eeq
provided that the operators transform as
\beq
\calO_i(x)\to |f'(x)|^{\Delta_i}\calO_i(x').
\eeq
Here $\Delta_i$ is a parameter characterizing the operator called its scaling dimension. All operators which are not primaries are derivatives of primaries of some finite order. They are called descendants.

One difference between CFT in $d=1$ and in $d>1$ is that the correlation functions generally depend on the point ordering as indicated. Notice that we can equivalently think of correlation functions as defined on a circle which is the conformal compactification of the real line. ${\rm SL}(2,\bR)$ transformations then preserve circular ordering. 
We may or may not assume parity invariance under $x\to -x$.
 
{\bf Axiom 2.} (Unitarity) There is an antilinear conjugation map defined on the primary operators: $\calO\to\calO^*$. This map has two properties: 
\begin{itemize}
\item (`Time-reversal') Correlation functions of conjugate operators in reversed positions are related by complex conjugation:
\beq
\langle \calO_n^*(-x_n)\ldots\calO_1^*(-x_1)\rangle = \langle \calO_1(x_1)\ldots\calO_n(x_n)\rangle ^*\,.
\eeq
Notice that this is different from parity invariance which would also relate correlation functions under $x\to-x$ but without complex conjugation.  

\item (Reflection positivity) The following reflection-symmetric linear combinations of $2n$-point correlation functions of primaries are non-negative:
\beq
\label{refl-pos}
\int d y\,\tilde g( y) \int d x\, g( x) \langle \calO_n^*(y_n)\ldots\calO_1^*(y_1) \calO_1(x_1)\ldots\calO_n(x_n)\rangle \ge 0\,,
\eeq
where $g(x)=g(x_1,\ldots,x_n)$ is an arbitrary function or distribution with support on 
\beq
0<x_1<x_2<\ldots<x_n
\eeq (away from coincident points to avoid singularities), and $\tilde g(y)$ is the reflected complex-conjugate function:
\beq
\tilde g(y_1,\ldots,y_n) = g(-y_1,\ldots,-y_n)^*\,.
\eeq

The physical meaning of this property is that we can consider a state
\beq
|\Psi\rangle = \int dx\, g(x) |\calO_1(x_1)\ldots\calO_n(x_n)|0\rangle\,.
\eeq
Then Eq.~\reef{refl-pos} can be rewritten as
\beq
 \langle\Psi|\Psi\rangle \ge 0\,.
 \eeq
 \end{itemize}
 
It is a simple consequence of reflection positivity (for $n=1$) that all primaries should satisfy the 1d unitarity bound $\Delta\ge 0$.\footnote{To exclude all $\Delta<0$, it suffices to consider $g(x)=\delta(x-x_0)+\alpha\, \delta'(x-x_0)$ for an appropriate $\alpha$.}

Let us pass to the basis of `hermitean' operators satisfying $\calO=\calO^*$. Hermitean primary operators are normalized so that their 2pt functions take the form:
\beq
\langle \calO_i(x_1)\calO_j(x_2) \rangle = \frac{\delta_{ij}}{|x_1-x_2|^{2\Delta_{\calO_i}}}.
\eeq
The ${\rm SL}(2,\bR)$ invariance implies that 3pt functions of primaries take the form:
\beq
\langle \calO_1(x_1)\calO_2(x_2) \calO_3(x_3)\rangle = \frac{C_{123}}{(x_2-x_1)^{h_{123}}(x_3-x_2)^{h_{231}}(x_3-x_1)^{h_{132}}}\qquad (x_1<x_2<x_3)\,,\\
\eeq
where $h_{ijk}=\D_i+\D_j-\D_k$. The 3pt function coefficients are in general complex (even for hermitean operators). In general, they depend on the circular ordering of the operators. It follows from the time-reversal property that $C_{321}=C_{123}^*$.

{\bf Axiom 3.} (OPE) $n$-point correlation functions of primaries can be reduced to $(n-1)$-point correlation functions using the OPE:
\beq
\calO_1(x_1) \calO_2(x_2) = \sum_k C_{12k} \frac{1}{(x_2-x_1)^{\Delta_1+\Delta_2-\Delta_k}}(\calO_k(x_1)+\ldots)
\eeq
where we assume the operator ordering $x_1<x_2$ in the l.h.s. The $\ldots$ in the r.h.s.~stands to the contributions of the descendants of $\calO_k$; they are fixed by the ${\rm SL}(2,\bR)$ invariance. For a continuous spectrum one should replace the sum by an integral.

The given axioms are sufficient to introduce the concept of conformal blocks, derive the representation \reef{CBdec} for the 4pt function of four identical hermitean primaries, and the positivity of the spectral density.\footnote{We have $p_k=C_{\phi\phi k}C_{k\phi\phi}=|C_{\phi\phi k}|^2$.}

Systems of correlation functions satisfying such axioms may arise in physics in a variety of ways.
One example are the generalized free theories discussed in the next section. One can also consider renormalization group flows starting from such generalized free theories, perturbing them by local operators. E.g.~one can consider the generalized free boson $\phi$ and perturb it by the operator $\phi^4$ (relevant for $\Delta<1/4$). This gives a flow to a fixed point which has ${\rm SL}(2,\bR)$ invariance \cite{Paulos:2015jfa}. This flow occurs in the studies of the critical point of the 1d long-range Ising model \cite{Fisher:1972zz} (see e.g.~\cite{Behan:2017dwr,Behan:2017emf} for recent work). Similar flows involving generalized free fermions were recently considered in connection with the Sachdev-Ye-Kitaev (SYK) model \cite{Gross:2017vhb}.\footnote{Note that the SYK model itself does not quite satisfy the axioms of 1d CFT, because of a small explicit breaking of the ${\rm SL}(2,\bR)$ invariance \cite{Maldacena:2016hyu}. However the nonlocal variation of the SYK model considered in \cite{Gross:2017vhb} does satisfy the 1d CFT axioms.}

A second way to get 1d CFTs is starting from higher-dimensional CFTs and introducing line defects into them (see e.g.~\cite{Billo:2016cpy} for a general discussion). For example one can take a 2d CFT in a half-plane (in this case the defect is a boundary), or two different 2d CFTs separated by a line interface. One can also consider line defects in 3d dimensional CFTs imposing nontrivial monodromy for the global symmetry when moving around the defect, as done for the 3d Ising model in \cite{Billo:2013jda,Gaiotto:2013nva} and for the $O(N)$ model in \cite{Soderberg:2017oaa}. Wilson line operators in conformal gauge theories also provide examples of line defects \cite{Giombi:2017cqn}. In all these examples, by taking OPE of higher-dimensional CFT operators with the defect one can define the local operators on the defect. Their correlators will satisfy axioms of 1d CFT.\footnote{We thank Subir Sachdev for bringing to our attention the following further example of 1d CFT arising in condensed matter physics \cite{Sachdev1,Sachdev2,Sachdev:2001ky}. Take a (2+1)-dimensional quantum antiferromagnet with spatially anisotropic couplings. Varying the anisotropicity parameter $\lambda$, one can reach a quantum critical point, which is a relativistic CFT in the $O(3)$ Wilson-Fisher universality class. Now introduce a magnetic impurity into the regular 2d lattice. The interaction between the impurity spin and the fluctuations of the N\'eel order parameter in the bulk antiferromagnet is relevant. It triggers an RG flow leading, at long time scales, to a fixed point which is a nontrivial 1d CFT living on the worldline of the impurity.}

A third way to get 1d CFT is to start with a general UV complete 2d QFT (which may be conformal or massive) and put it in a classical curved background with AdS${}_2$ geometry. In this setup 1d CFT lives on the boundary of the AdS space and provides a dual description of bulk physics (see \cite{Paulos:2016fap} for a recent use of this observation).

One necessary property of the 1d CFTs is that they are nonlocal, in that they do not have an analogue of a local conserved stress tensor operator, unlike local CFTs possible in $d\ge 2$. This is formally obvious already from the fact that any conserved operator in $d=1$ is a constant. Physically the nonlocality of these theories can be traced back to their construction: 
\begin{itemize}
\item
Generalized free theories are nonlocal because they can be defined by a gaussian non-local action (see e.g.~the detailed discussion in \cite{Paulos:2015jfa}). Flows originating from such theories are thus also expected to be nonlocal. 

\item
1d CFTs arising as boundaries or defects of higher-dimensional CFTs are also nonlocal when viewed in isolation. Their local description is impossible without taking into account the higher-dimensional bulk. 

\item 1d CFTs arising from QFTs in AdS are nonlocal because we are considering a classical, non-fluctuating, AdS background. This is to be contrasted with the more standard AdS/CFT setting in higher dimensions, where the gravitational background is allowed to fluctuate, the graviton field describing these small fluctuations being dual to the local stress tensor operator on the boundary.
\end{itemize}

{Finally, we note that 1d CFTs have been previously studied via numerical bootstrap in \cite{Gaiotto:2013nva,El-Showk:2016mxr,Paulos:2016fap}, and by analytic bootstrap techniques in \cite{Mazac:2016qev,Hogervorst:2017sfd,Rychkov:2017tpc}. }

\section{Generalized free theories}
\label{GFF}

The two simplest unitary 1d CFTs are the generalized free boson and fermion. These are gaussian theories, in the sense that $n$-point correlation functions are expressed in terms of the 2pt function via Wick's theorem. The bosonic ($+$) and fermionic ($-$) 4pt functions are given by:
\begin{multline}
\label{eq:GFF}
\langle \phi(x_1) \phi(x_2)\phi(x_3)\phi(x_4)\rangle
= \langle \phi(x_1) \phi(x_2)\rangle\langle\phi(x_3)\phi(x_4)\rangle
\\+
\langle \phi(x_1) \phi(x_4)\rangle\langle\phi(x_2)\phi(x_3)\rangle
\pm \langle \phi(x_1) \phi(x_3)\rangle\langle\phi(x_2)\phi(x_4)\rangle 
\end{multline}
where the points are ordered as $x_1<x_2<x_3<x_4$ and
\beq
 \langle \phi(x_1) \phi(x_2)\rangle = (x_2-x_1)^{-2\Delta_\phi}.
\eeq
The corresponding functions $\calG(z)$ are given by
\be
\calG(z) = 1 \pm z^{2\Df} +\left(\frac{z}{1-z}\right)^{2\Df}.
\ee
The exact conformal block decomposition is known. In addition to the unit operator, it contains operators of dimension ($i\ge 0$)
\beq
\Delta_i=\begin{cases} 2\Df+2 i & \text{(boson)}\\
2\Df+2 i+1 & \text{(fermion)}\,,
\end{cases}
\eeq
while the squared OPE coefficients are given by (see \cite{Gaiotto:2013nva}, Eqs.~(4.14-15), and \cite{Hogervorst:2017sfd}, section 2.3.1)
\beq
p_i=\frac{(2\Df)_i}{4^i}\times
\begin{cases}  
\dfrac{2 (2\Df)_{2i}}{(2i)! (2\Df+i-1/2)_{i}} & \text{(boson)}\\
\dfrac{(2\Df)_{2i+1}}{(2i+1)! (2\Df+i+1/2)_{i}}& \text{(fermion)}.
\end{cases}
\eeq
Expressing the coefficients in terms of the operator dimension, we have in both cases
\beq
p(\Delta)\approx \dfrac{8\sqrt{\pi}}{\Gamma({2\Df})^2}\D^{4\Df-3/2}4^{-\D}\qquad(\Delta\gg 1).
\eeq
Recalling that the operator dimensions increase in step of 2, we see that this asymptotics is in agreement with \reef{ptilde}, \reef{gammaA}.

\section{$I\sim J$}
\label{appIJ}

Here we will show \reef{IJ}. It is convenient to change the integration variable to $y$:
\beq
I = \int_0^\infty \frac{dy}{\sqrt{y(y+4)}} (1+\kappa y)^{-\D}
\eeq
and similarly for $J$. Let us split this integral into two parts: 
\beq
I=\int_0^{y_*}+\int_{y_*}^\infty=I_1+I_2\,,
\eeq
and similarly for $J$. We will fix an $\eps>0$ and choose $y_*=1/(\kappa \Delta^{1-\eps})$. 

We will show that, uniformly in $\kappa$,
\beq
\label{I2I1}
I_2/I_1, J_2/J_1 = O(e^{-\D^{\eps}})
\eeq
and 
\beq
\label{I1J1}
I_1/J_1 = 1+O(1/\D^{1-2\eps})\,.
\eeq
These two facts clearly imply \reef{IJ}, with $a=1-2\eps$.

Let us show \reef{I2I1}. Using the monotonicity of the integration measure, we have
\begin{align}
I_1&\ge \frac{1}{\sqrt{y_*(y_*+4)}}\int_0^{y_*} dy\,(1+\kappa y)^{-\D},\nn\\
I_2&\le \frac{1}{\sqrt{y_*(y_*+4)}}\int_{y_*}^{\infty} dy\, (1+\kappa y)^{-\D}.\label{monot}
\end{align}
Doing the integrals in the r.h.s.~and taking the ratio, the $y_*$-dependent prefactors cancel and we get
\beq
I_2/I_1 \le  
\frac{(1+1/\Delta^{1-\eps})^{1-\D}}
{1-(1+1/\Delta^{1-\eps})^{1-\Delta}},
\eeq
from where \reef{I2I1} follows. A similar argument works for $J_2/J_1$. 

On the other hand, \reef{I1J1} follows from the fact that the ratio of integrands is everywhere close to 1 on the corresponding interval:
\beq
\log \frac{(1+\kappa y)^{-\D}}{e^{-\kappa y \D}} = [-\log(1+\kappa y) +\kappa y]\D=O((\kappa y_*)^2\Delta)=O(1/\D^{1-2\eps})\,.
\eeq

\section{Tauberian theorem without exact prefactor}

\label{sec:simple}

Here we will give a simple argument that condition \reef{w1}, which we copy here for convenience,
\beq
\int_0^\infty d\D\,q(\D) w_1(\Delta/Y) \sim Y^\gamma \qquad (Y\to \infty)
\label{w1copy}
\eeq
 implies that\footnote{The lower bound is slightly better than $C_2 Y^\gamma/\log Y$ proved in \cite{Fitzpatrick:2012yx}, appendix B.3.} ($Q(Y)= \int_0^Y d\D\, q(\Delta)$)
\beq
C_2 Y^\gamma\le Q(Y)  \le C_1 Y^\gamma\qquad(Y\gg 1)\,.
\label{simpleTaub}
\eeq
For this argument we will assume that $w_1$ is a positive function which is differentiable and monotonically decreasing. There will be also a few extra conditions which will introduce when we need them.
For $w_1$ as in \reef{notation} all these conditions will be satisfied.

Since $w_1$ is monotonically decreasing, we have
\beq
\int_0^\infty d\D\,q(\D) w_1(\Delta/Y) \ge w_1(1) Q(Y)\,,
\eeq
From here the r.h.s.~inequality in \reef{simpleTaub} for $Y\gg 1$ follows. Increasing the constant $C_1$ somewhat if needed, we can achieve that this inequality is true for all $Y>0$. This is possible because $q(\Delta)$, and hence $Q(\Delta)$, vanishes for $\Delta\le \Delta_0$.

Now let us show the other direction. We rewrite integrating by parts
\begin{align}
\int_0^\infty d\D\,q(\D) w_1(\Delta/Y) &= - Y^{-1} \int_0^\infty d\D\, Q(\Delta) w_1'(\Delta/Y)\nn\\
 &= \int_0^\infty \frac{d\D}\D\, Q(\Delta) g(\Delta/Y) \nn\\
 &= \int_0^\infty \frac{d t}t\, Q(Y t) g(t) \,,
 \label{rewrite}
\end{align}
where we denoted $g(t)=-t w_1'(t)$ (a positive function by assumption). We dropped the boundary terms in the first line of this equation. The $t=0$ boundary term vanishes because $Q(\Delta)=0$ for small $\Delta$. Using the already proved r.h.s.~inequality in \reef{simpleTaub}, the $t=\infty$ boundary term vanishes if
\beq
t^\gamma w_1'(t) \to 0 \qquad(t\to\infty).
\eeq
We will assume that this is also satisfied. 

Let us split the last integral in \reef{rewrite} into three parts 
$\int_0^{t_1}+\int_{t_1}^{t_2}+\int_{t_2}^\infty$. In the first and the last integral we bound the integrand using the already proved upper bound in \reef{simpleTaub} (valid for any $Y$ as mentioned), which gives
\beq
\label{tailsEst}
\int_0^{t_1}+\int_{t_2}^\infty \le B_1 Y^\gamma,\qquad B_1 = C_1 \left(\int_0^{t_1}+\int_{t_2}^\infty\right) dt\,t^{\gamma-1} g(t) .
\eeq
We will assume that $\int_0^\infty dt\,t^{\gamma-1} g(t)$ converges (this is satisfied for our $w_1$ taking into account $\gamma>0$). Given this, we can pick $t_1$ and $t_2$ so that $B_1<1$. Then the remaining part of the integral, $\int_{t_1}^{t_2}$, should be asymptotically larger than $(1-B_1)Y^\gamma$. On the other hand we have
\beq
\int_{t_1}^{t_2} \frac{d t}t\, Q(Y t) g(t) \le B_2\, Q(Y t_2) ,\qquad B_2 = \int_{t_1}^{t_2} \frac{d t}t\, g(t).
\eeq
This implies the l.h.s.~inequality in \reef{simpleTaub} with $C_2 = (1-B_1)/(B_2 t_2^\gamma)$.

\section{Karamata's argument} 
\label{Karamata}
The Hardy-Littlewood theorem can be given an elementary proof using a beautiful argument due to Karamata \cite{Titchmarsh}. The argument runs as follows. Consider for simplicity the case $\gamma=1$ (the general case being similar). Suppose that
\beq
\int_0^\infty d\Delta\, r(\Delta)\, f(\Delta\, t)\sim t^{-1} \qquad(t\to0),
\eeq
where we denoted $f(t)=e^{-t}$. Then clearly
\beq
\label{n+1}
 \int_0^\infty d\Delta\, r(\Delta)\, f(\Delta (n+1) t)\sim t^{-1} \frac{1}{n+1}\qquad(t\to0).
\eeq
The r.h.s. can be written as 
\beq
t^{-1}\int_0^\infty dy\,f((n+1)y)\left/\int_0^\infty dy\,f(y)\right..
\eeq
Taking finite linear combinations of \reef{n+1} we obtain
\beq
\label{Pint}
\int_0^\infty d\Delta\, r(\Delta)\, P(\Delta\, t)\sim t^{-1} \int_0^\infty dy\,P(y)\left/\int_0^\infty dy\,f(y)\qquad(t\to0)\right.
\eeq
for any $P(y)$ which is of the form
\beq
\label{P}
P(y)=\sum_{n=0}^N c_n f((n+1)y).
\eeq
Now suppose that we can find functions $P_+(y)$ and $P_-(y)$ which are of the form \reef{P}
and such that \emph{Karamata's conditions} hold:
\beq
\label{comp0}
\tag{K1}
P_-(y)\le \Theta(y\in[0,1]) \le P_+(y)\qquad (0\le y\le \infty),
\eeq
while the integrals
\beq
\label{comp1}
\tag{K2}
\left|\int_0^\infty P_\pm(y) dy -1\right |\le \eps
\eeq
with $\eps$ arbitrarily small.

Since the spectral density is non-negative, we can integrate \reef{comp0} and obtain:
\beq
\int_0^\infty d\Delta\, r(\Delta)\, P_-(\Delta\, t)\le \int_0^\infty d\Delta\, r(\Delta)\, \Theta(\Delta\, t \in[0,1]) \le
\int_0^\infty d\Delta\, r(\Delta)\, P_+(\Delta\, t).
\eeq
By \reef{Pint}, the sides of this inequality behave as $\sim t^{-1}(1+O(\eps))$. Since $\eps$ is arbitrary, the middle function must behave as $\sim t^{-1}$. This middle function is exactly 
\beq
\int_0^{1/t} d\Delta\, r(\Delta),
\eeq
and so we proved the theorem.

We still have to show that \reef{comp0}, \reef{comp1} can be satisfied. For $f(t)=e^{-t}$ this is easy to show introducing a new variable $x=e^{-t}\in[0,1]$. Then finite sums \reef{P} are polynomials and functions $P_\pm$ can be shown to exist applying the Weierstrass approximation theorem. See \cite{Titchmarsh} for details.

One may be wondering if the conformal bootstrap tauberian theorem can be proved by Karamata's argument \cite{Fitzpatrick:2014vua}. For this one would need to show that conditions \reef{comp0}, \reef{comp1} can be satisfied for $f(y)=K_0(y)$. We do not know if this can be done. The Weierstrass theorem, or its generalization the Stone-Weierstrass theorem, require that the approximating set of functions form an algebra under pointwise multiplication. 
This condition is satisfied by the polynomials but not by the linear combinations of $K_0(n y)$.

Ref.~\cite{Fitzpatrick:2014vua}, appendix F.2, note 9, tried to reduce Karamata's conditions for $f(y)=K_0(y)$ to those for $f(y)=e^{-y}$. They proposed to use an integral transforms to represent $e^{-y}$ as a multiplicative convolution of $K_0(y)$. However, this seems impossible because the Mellin transform of $K_0(y)$ decreases faster at infinity (these Mellin transforms are identical with the Fourier transforms \reef{MW1} and \reef{MW1HL}). In any case, even if such an integral representation exists, it will involve a continuum of dilatations of $K_0(y)$, while in \reef{P} we need a finite linear combination.\footnote{We thank Liam Fitzpatrick for the correspondence related to these points.}

So, straightforward adaptation of Karamata's argument does not seem to work for the conformal bootstrap tauberian theorem. Still, one can see the proof we presented in section \ref{proof} as ``Karamata's argument on steroids", as there is a clear analogy between Karamata's conditions and our (\ref{squeeze}), (\ref{squeeze1}). 
That in our proof we passed from dilatations to translations is just a matter of convenience. On the other hand the important extra ingredients were a systematic and justified use of convolution, as opposed to finite linear combinations, and the control over the Fourier transform at infinity needed to find convolution representations.

\section{Connection to the lightcone bootstrap for $d>1$}
\label{d>1}
The initial motivation for considering a tauberian theorem of the kind we proved in this paper came from analyzing the conformal bootstrap equations near the lightcone \cite{Fitzpatrick:2012yx,Komargodski:2012ek}. Let us review this connection. 

\begin{figure}
\centering
\includegraphics[width=0.5\textwidth]{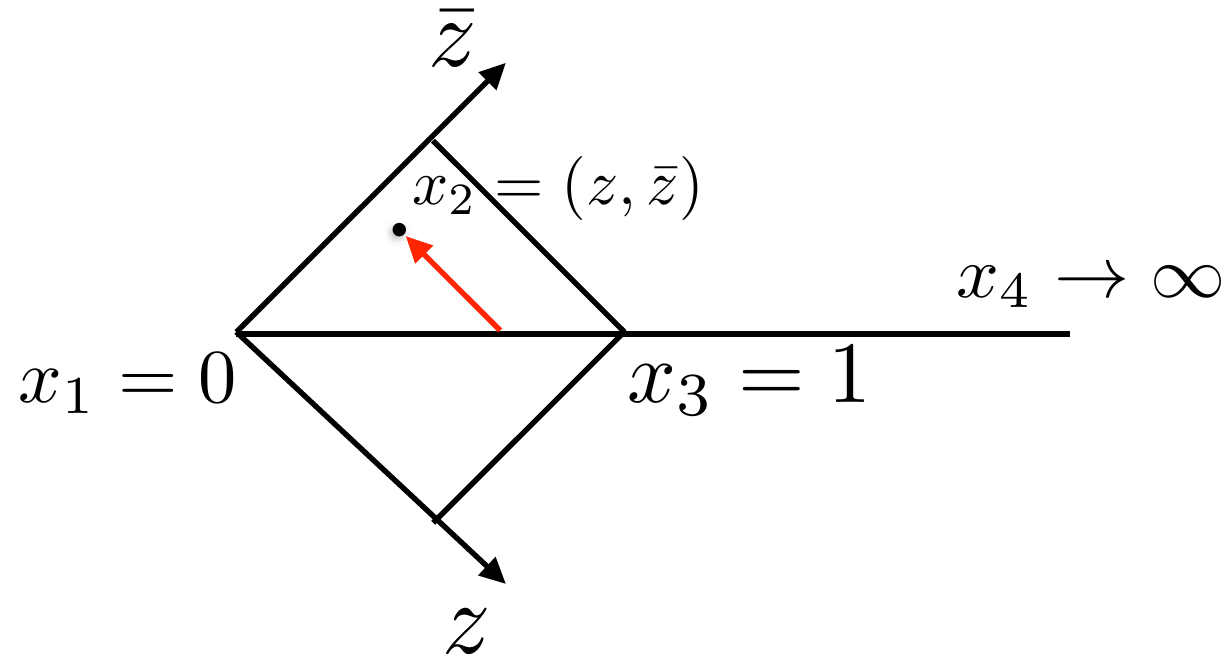}
\caption{The causal diamond where the Minkowski 4pt function is nonsingular. The 4pt function is real and positive in the diamond, symmetric with respect to the diagonal $z=\bar z$. The function $\calG(z,\bar z)$ decreases from the diagonal in the direction shown by the arrow.}
\label{diamond}
\end{figure}

In a unitary CFT in $d>1$ dimensions, consider a 4pt function of four identical scalar primaries
\beq
\langle \phi(0) \phi(z,\bar z) \phi(1) \phi(\infty)\rangle = {(z \bar z)^{-\Df}} \calG(z,\bar z).
\eeq
Here we are placing four operators in a two-dimensional plane (which is a subspace of the full $d$-dimensional space if $d>2$), putting them at four points in this plane as indicated. Crucially, we will be considering this plane being of Minkowski signature, analytically continuing correlators from the Euclidean. For the Minkowski signature the coordinates $z$ and $\bar z$ are two independent real coordinates, and the 4pt function is real and smooth (real-analytic) when the operator $\phi(z,\bar z)$ is in the causal diamond $0<z,\bar z<1$ limited by the null rays emanating from the operators $\phi(0)$ and $\phi(1)$, see figure \ref{diamond}.

The 4pt function can be expanded into conformal blocks $g_{\tau,\ell}(z,\bar z)$ of primaries labeled by their spin $\ell$ and twist $\tau=\Delta-\ell$:
\beq
\label{d>1exp}
\calG(z,\bar z)=1+\sum_{\ell,\tau} P_{\tau,\ell}\, g_{\tau,\ell}(z,\bar z)\,.
\eeq
Conformal blocks themselves have in the causal diamond a convergent power series expansion of the form
\beq
\label{d>1CB}
g_{\tau,\ell}(z,\bar z) = (z\bar z)^{\tau/2}\sum_{m,n=0}^\infty a_{m,n} z^{m} {\bar z}^{n}.
\eeq
In unitary theories coefficients $P_{\tau,\ell}$ are real and positive. For primaries above the unitarity bounds all coefficients $a_{m,n}$ are also real and positive. It follows that all terms in the expansions \reef{d>1exp} and \reef{d>1CB} are real and positive, and so the expansion must converge to the 4pt function (no possibility for any cancellation).

Moreover we can derive a bound on the 4pt function in the diamond in terms of its value on the diagonal. Let us decrease $z$ starting from a point on the diagonal $z=\bar z \in (0,1)$. In unitary theories in $d\ge 2$ we have unitarity bounds:
 \beq
 \tau \ge \tau_{0}=(d-2)/2\ge 0.
 \eeq
It follows that as we decrease $z$ every term in \reef{d>1exp} except for the first term is decreasing at least as fast as $z^{\rm \tau_0}$. We thus have:
\beq
\label{diamond-bound}
1\le \calG(z,\bar z)\le \calG(\bar z,\bar z) \qquad (0<z\le \bar z).
\eeq

In what follows we assume that our theory has a twist gap: all operators but the unity have a positive twist $\tau\ge \tau_{\min}>0$. This assumption follows automatically from the unitarity bounds in $d>2$. It is not satisfied in local CFTs in $d=2$ where the stress tensor and other conserved currents have twist zero, but it may be satisfied in nonlocal two-dimensional CFTs. By the same argument, the bound \reef{diamond-bound} can then be strengthened as follows
\beq
\label{diamond-bound1}
1\le \calG(z,\bar z)\le 1+(z/\bar z)^{\tau_{\min}/2} [\calG(\bar z,\bar z)-1] \qquad (0<z\le \bar z).
\eeq

Consider now the crossing equation for the 4pt function, which takes the form
\beq
\label{eq:cross}
(z\bar z)^{-{\Df}} \calG(z,\bar z)=[(1-z)(1-\bar z)]^{-\Df} \calG(1-z,1-\bar z).
\eeq

First of all let us derive an auxiliary result by considering the crossing equation on the diagonal and taking the limit $z=\bar z\to 1$. We can use the OPE in the r.h.s., which is dominated by the unit operator, and conclude
\beq
\calG(\bar z, \bar z) \sim (1-\bar z)^{-2\Df}\qquad (\bar z\to 1).
\eeq
In particular we have 
\beq
\label{eq:aux}
\calG(\bar z, \bar z) =O( (1-\bar z)^{-2\Df})\qquad (0<\bar z<1).
\eeq

\begin{figure}
\centering
\includegraphics[width=0.17\textwidth]{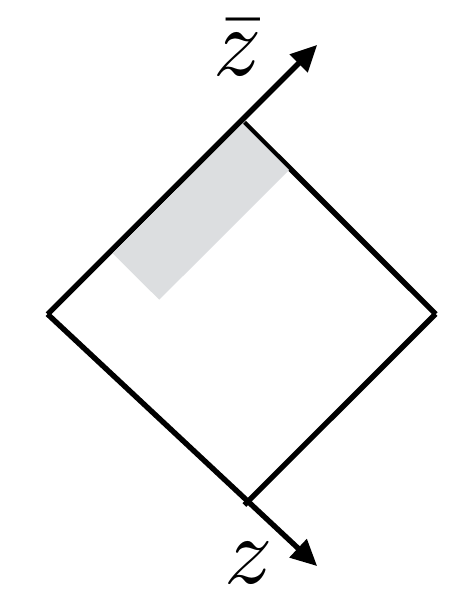}
\caption{The region in which we are analyzing crossing.}
\label{limit}
\end{figure}

Next we would like to consider the crossing equation away from the diagonal. We will be interested in the limit $z\to0$ while $\bar z$ will belong to some fixed range close to 1 (figure \ref{limit}). In this region, taking into account \reef{eq:aux}, we can rewrite the bound \reef{diamond-bound1} as
\beq
\label{eq:errorterm}
\calG(z,\bar z) = 1+O\left(z^{\tau_{\min}/2}(1-\bar z)^{-2\Df}\right).
\eeq
So we see that $\calG(z,\bar z)\to 1$ as $z\to 0$ although approach becomes slower and slower as $\bar z\to 1$.\footnote{In Mean Field Theory (see below), by inspection of the exact four point function, one can see that we can replace $(1-\bar z)^{-2\Df}$ by $(1-\bar z)^{-\Df}$ in the error term. We do not know if such an improvement is possible for general CFTs. We thank Amit Sever for raising this question.}

In any case we see that the l.h.s.~of \reef{eq:cross} behaves as 
\beq
\label{eq:as-reg}
(z\bar z)^{-{\Df}}\times [1+O\left(z^{\tau_{\min}/2}(1-\bar z)^{-2\Df})\right)]
\eeq
in the considered limit. The leading singular behavior is $const.z^{-{\Df}}$. We are interested in how this singular behavior can arise from the crossed channel. 

The key insight of \cite{Fitzpatrick:2012yx,Komargodski:2012ek} is that this can only happen due to cooperative behavior of many primaries of high spin and approximately constant twist. The individual conformal blocks in the crossed channel behave for $z\ll 1$, $l\gg 1$, $\bar z$ fixed as \cite{Fitzpatrick:2012yx}
\beq
\label{eq:CBas-d}
g_{\tau,\ell}(1-z,1-\bar z) = G_{\ell}(1-z) (1-\bar z)^{\tau/2} F(\tau,\bar z)\times[1+O(1/\sqrt{\ell},\sqrt{z})].
\eeq
Here $G_{\ell}(z)$ are the 1d conformal blocks defined in \reef{CBdec}, for $\Delta=\ell$. The function $F$ is a $d$-specific function which is analytic, regular and positive at $\bar z=1$. 

Neglecting the error terms in \reef{eq:as-reg} and \reef{eq:CBas-d}, the crossing equation becomes
\beq
\label{eq:becomes}
z^{-{\Df}}\sim \sum_{\ell,\tau} P_{\tau,\ell}\, G_{\ell}(1-z) \times (1-\bar z)^{\tau/2-\Df} \bar z^{\Df} F(\tau,\bar z).
\eeq
This equation is similar to the problem \reef{CBdec}, \reef{zto1} which we studied in this paper, but it's also rather more nontrivial. Notice that we are dealing with an asymptotics in $z$ where coefficients are functions dependent on another parameter $\bar z$. The asymptotics should be valid for any $\bar z$ close to one, but the range of $z$ when it kicks in becomes smaller and smaller as $\bar z\to 1$; it can be estimated from \reef{eq:errorterm} as
\beq
z\ll (1-\bar z)^{4\Df/\tau_{\min}}.
\eeq
This is a significant complication in analyzing the consequences of \reef{eq:becomes}.

If one ignores this complication and proceeds naively, one can try to argue as follows. The factor $(1-\bar z)^{\tau/2-\Df}$ in $\bar z$ dependent coefficients in the r.h.s.~is non-analytic in the limit $\bar z\to 1$, unless operators have twists
\beq
\label{eq:MFTtwists}
\tau = 2\Df+2n\qquad(n=0,1,2\ldots).
\eeq
In a generic CFT there will be small deviations from these twists, but let us assume that these deviations can be neglected in the limit $l\to \infty$. One then plugs twists \reef{eq:MFTtwists} into \reef{eq:becomes}, takes the limit $\bar z\to 1$
and requires that the equation should be satisfied order by order in $(1-\bar z)$. 
For example from the $(1-\bar z)^0$ term one gets that the $n=0$ series must satisfy:
\beq
\label{eq:n=0}
z^{-{\Df}}\sim F(2\Df, 1) \sum_{\ell} P_{2\Df,\ell}\, G_{\ell}(1-z) .
\eeq
The $n=1$ series must satisfy a similar equation needed to satisfy the equation at the order $(1-\bar z)^1$ etc.

We would like to emphasize that imposing the equation order by order in $(1-\bar z)$ appears to us a nontrivial assumption. It would be nice to justify it rigorously. A key difficulty in doing this is that the $z\to 0$ asymptotics is not valid uniformly for $\bar z$ near 1. {This problem appears to us more difficult than the one we solved in this paper, and it deserves further study.}

Ref.~\cite{Fitzpatrick:2012yx} tried to formalize the above way of reasoning as follows. They proposed to define a spectral density in the twist space
\beq
\label{eq:rhodef}
\rho(\sigma) = \lim _{z\to 0}z^{\Df}  \sum_{\tau,\ell} P_{\tau,\ell}\, G_{\ell}(1-z) \delta(\tau-\sigma).
\eeq
The limit should be understood in the sense of linear functionals, integrated against a continuous function $f(\tau)$. It is an open problem to show that the limit exists in any CFT.\footnote{The existence of the limit after integration with an  $f(\tau)$ is not obvious as $z^{\Df} G_l(1-z)$ is not monotonic as $z\to 0$. Therefore, while the function under the limit sign would remain bounded as $z\to0$ for any bounded $f(\tau)$, it may a priori oscillate. Ref.~\cite{Fitzpatrick:2012yx} tacitly assumes that the limit exists (see their Eq.~(91)) but does not provide a proof. However, to appeal to the Riesz Representation Theorem as they subsequently do, it is definitely necessary to first ascertain that the limit exists for a class of $f(\tau)$. We thank David Simmons-Duffin and Jared Kaplan for the comments on this point.}

Assuming that the limit exists, the spectral density must satisfy the equation
\beq
\label{eq-for-rho}
1=\int_{\tau_{\min}}^\infty d\sigma\, \rho(\sigma) (1-\bar z)^{\sigma/2-\Df} \bar z^{\Df} F(\sigma,\bar z)
\eeq
for any $\bar z$ close to 1. To solve this equation, one considers the generalized free scalar field in $d$ dimensions, also known as Mean Field Theory (MFT). In this theory the 4pt function is given in $d>1$ by \reef{eq:GFF} with the $+$ sign, and the conformal block decomposition can be found exactly. The only operators present in the OPE are those of twists exactly \reef{eq:MFTtwists}. The OPE coefficients can be found exactly, and expanding them at large spin and performing the sum in \reef{eq:rhodef} one finds that in this particular case the limit does exist and is given by a sum of delta-functions
\beq
\rho(\sigma) = \sum_{n=0}^\infty P^{\rm{MFT}}_{2\Df+2n}\delta(\sigma - (2\Df+2n))\,,
\eeq
with $P^{\rm{MFT}}_{2\Df+2n}$ given in Ref.~\cite{Fitzpatrick:2012yx}, Eq.~(25).
This, then, is one solution of Eq.~\reef{eq-for-rho}. One then argues that this solution is unique.

We thus obtain the following generalization of Eq.~\reef{eq:n=0}:
\beq
\label{eq:n-gen}
\sum_{\ell} P_{2\Df+2n,\ell}\, G_{\ell}(1-z) \sim P^{\rm{MFT}}_{2\Df+2n} z^{-{\Df}}\qquad(n=0,1,2\ldots),
\eeq
where $P_{2\Df+2n,\ell}$ denotes OPE coefficients of a series of operators approaching twist $2\Df+2n$ in the large spin limit.

Eqs.~\reef{eq:n=0}, \reef{eq:n-gen} are of the same basic form as the problem \reef{CBdec}, \reef{zto1} which we studied in this paper, with a minor redefinition of parameters and with a restriction that $\Delta$'s must be integers. Using our results, we can rigorously show that Eq.~\reef{eq:n-gen} implies an asymptotic for the integrated OPE coefficients of each series of operators. These are the asymptotics conjectured in \cite{Fitzpatrick:2012yx}, Eq.~(32).

It would be interesting to put on more solid ground other parts of the argument in \cite{Fitzpatrick:2012yx}. This concerns in particular the existence of the limit for $\rho(\sigma)$ in \reef{eq:rhodef}. Also the arguments in \cite{Fitzpatrick:2012yx}, section 2.3 (and analogous arguments in \cite{Komargodski:2012ek}), need further justification. There, the subleading terms in the $z\to0$ and $\bar z\to 1$ asymptotics are conjectured to imply subleading powerlaw asymptotics of OPE coefficients and twists as $\ell\to\infty$. We are not aware of any mathematical theorem which can be used to show this rigorously in full generality.

In general, it's subtle to control subleading terms in tauberian theorems \cite{Korevaar}. In mathematics literature, it's known that the ability to go out to the complex plane is very useful in this task, see \cite{Korevaar}, Chapter III. In our work here, as well as in earlier arguments in \cite{Fitzpatrick:2012yx,Komargodski:2012ek}, only real analysis methods were used. A different approach to the lightcone bootstrap was recently proposed in Ref.~\cite{Caron-Huot:2017vep}, which relies crucially on the analytic structure of the CFT 4pt function for complex values of $z$, $\bar z$. This approach leads to an explicit inversion formula which defines an analytic function of $\Delta$, $\ell$, whose poles encode the physical spectrum of the CFT. It is possible that further investigations based on this formula will bring clarity to the problems discussed in this section. 

\small \bibliography{tauberian.bib}

\bibliographystyle{utphys.bst}

\end{document}